\documentclass[astrosymb]{aastex631}

\graphicspath{{./}{figures/}}
\usepackage[figuresright]{rotating}
\usepackage{color}
\usepackage{amsmath}
\usepackage{epstopdf}
\usepackage{threeparttable}
\usepackage{booktabs}
\begin{document}

\title{The formation of the stripped envelope type $\rm \uppercase\expandafter{\romannumeral2}$b Supernova progenitors: Rotation, Metallicity and Overshooting}

\author{Gang Long}
\affiliation{College of Physics, Guizhou University, Guiyang city, Guizhou Province, 550025, P.R. China}

\author{Hanfeng Song}
\affiliation{College of Physics, Guizhou University, Guiyang city, Guizhou Province, 550025, P.R. China}
\affiliation{Geneva Observatory, Geneva University, CH-1290 Sauverny,Switzerland}

\author{Georges Meynet}
\affiliation{Geneva Observatory, Geneva University, CH-1290 Sauverny,Switzerland}

\author{Andre Maeder}
\affiliation{Geneva Observatory, Geneva University, CH-1290 Sauverny,Switzerland}

\author{Ruiyu Zhang}
\affiliation{College of Physics, Henan Normal University, Xinxiang, Henan Province, 453007, P.R. China}

\author{Ying Qin}
\affiliation{Department of Physics, Anhui Normal University, Wuhu city, Anhui Province, 241000, P.R. China}

\author{Sylvia Ekstr\"omt}
\affiliation{Geneva Observatory, Geneva University, CH-1290 Sauverny,Switzerland}

\author{Cyril Georgy}
\affiliation{Geneva Observatory, Geneva University, CH-1290 Sauverny,Switzerland}

\author{Liuyan Zhao}
\affiliation{College of Physics, Guizhou University, Guiyang city, Guizhou Province, 550025, P.R. China}

\correspondingauthor{Hanfeng Song; Georges Meynet; Andre Maeder}
\email{hfsong@gzu.edu.cn;Georges.Meynet@unige.ch; Andre.Maeder@unige.ch}



\begin{abstract}

Type $\rm \uppercase\expandafter{\romannumeral2}$b supernovae are believed to originate from core-collapse progenitors having kept only a very thin hydrogen envelope. We aim to explore how some physical factors, such as rotation, metallicity,  overshooting, and the initial orbital period in binaries, significantly affect the Roche lobe overflow and the formation of type $\rm \uppercase\expandafter{\romannumeral2}$b supernovae. It is found that binaries are the main channel that capable of producing type type $\rm \uppercase\expandafter{\romannumeral2}$b supernovae progenitors in the mass range for initial masses below 20 $M_{\odot}$. The formation of type $\rm \uppercase\expandafter{\romannumeral2}$b supernova progenitors is extremely sensitive to the initial orbital period. A less massive hydrogen indicates smaller radius and a higher effective temperatures, and vice versa. Binary systems with initial periods between 300 and 720 days produce type $\rm \uppercase\expandafter{\romannumeral2}$b progenitors that are a red supergiant. Those with an initial period between 50 and 300 days produce yellow supergiant progenitors and those with initial periods shorter than 50 days, blue supergiant progenitors. Both rapid rotation and larger overshooting can enlarge the carbon-oxygen core mass and lead to higher core temperature and lower central density at the pre-collapse phase. They are also beneficial to surface nitrogen enrichment but restrict the efficiency of the first dredge-up. SN $\rm \uppercase\expandafter{\romannumeral2}$b progenitors with low metallicity have smaller hydrogen envelope masses and radii than the high metallicity counterparts. Ultra-stripped binary models have systematically higher core mass fraction $\rm ^{12}C$ left, which has important influence on the compactness of type $\rm \uppercase\expandafter{\romannumeral2}$b progenitors.

\end{abstract}


\section{Introduction}

Massive stars explode as core-collapse with varying amounts of hydrogen in their envelopes.
Supernovae are classified as Type $\rm \uppercase\expandafter{\romannumeral1}$ and Type $\rm \uppercase\expandafter{\romannumeral2}$ in terms of the absence and presence of hydrogen lines in the spectrum and further subdivided into $\rm \uppercase\expandafter{\romannumeral1}$a, $\rm \uppercase\expandafter{\romannumeral1}$b, $\rm \uppercase\expandafter{\romannumeral1}$c, $\rm \uppercase\expandafter{\romannumeral2}$P, $\rm \uppercase\expandafter{\romannumeral2}$L, $\rm \uppercase\expandafter{\romannumeral2}$b, and IIn \citep[e.g.,][]{Branch1991,Filippenko1991}. The spectra of type $\rm \uppercase\expandafter{\romannumeral1}$ core-collapse supernovae show an absence of hydrogen lines.
The presence of strong Si lines and He lines defines Type $\rm \uppercase\expandafter{\romannumeral1}$a and Type $\rm \uppercase\expandafter{\romannumeral1}$b, respectively, while Type $\rm \uppercase\expandafter{\romannumeral1}$c does not display any hydrogen and helium features in their spectra. There are a few signatures of hydrogen in the spectra of type $\rm \uppercase\expandafter{\romannumeral1}$b.

It has been suggested that the diversity of SNe $\rm \uppercase\expandafter{\romannumeral2}$ originates from different main-sequence mass ranges of the progenitor , i.e., SNe $\rm \uppercase\expandafter{\romannumeral2}$L from about 7-10$M_{\odot}$, SNe $\rm \uppercase\expandafter{\romannumeral2}$P above 10$M_{\odot}$, and Type $\rm \uppercase\expandafter{\romannumeral1}$b/$\rm \uppercase\expandafter{\romannumeral1}$c supernovae (SNe $\rm \uppercase\expandafter{\romannumeral1}$b/$\rm \uppercase\expandafter{\romannumeral1}$c) originating from He stars of different mass ranges in binary systems \citep{Nomoto1990}.
However, the exact connection of these types to their progenitors has been a controversial issue.

In this paper, we focus on stripped-envelope SNe, e.g., SNe of Type
$\rm \uppercase\expandafter{\romannumeral2}$b, Ib and Ic. Type $\rm \uppercase\expandafter{\romannumeral2}$b SNe, which
are considered as a transitional class, showing clear hydrogen signatures in
their early spectra. However, these signatures gradually disappear, over the
period of 30-90 days after the explosions, after which the spectra become
virtually indistinguishable from Type $\rm \uppercase\expandafter{\romannumeral1}$b SNe. Possibly the simplest explanation for these observational differences
among the stripped envelope (SE) SNe subclasses is that different amounts of the
helium/hydrogen envelopes have been stripped from the star prior to the
SN explosions. In this channel, type $\rm \uppercase\expandafter{\romannumeral1}$c SNe are stripped the most whereas $\rm \uppercase\expandafter{\romannumeral2}$b
the least.

The light curve of SN $\rm \uppercase\expandafter{\romannumeral2}$b can show an early fireball phase but is powered near maximum and beyond by radioactive decay. The thin envelope model has been confirmed by spectral variations which show growing features of helium and oxygen.
The light curve of SN $\rm \uppercase\expandafter{\romannumeral2}$b is significantly different from the previously known light curve of SNe II. It is obvious that the peculiar light curve of SN $\rm \uppercase\expandafter{\romannumeral2}$b cannot be explained by the explosion of an ordinary red supergiant with a massive hydrogen-rich envelope, which produces a light curve of SN II-P. The light curve of SN $\rm \uppercase\expandafter{\romannumeral2}$b can be understood as the explosion of a red-supergiant whose hydrogen-rich envelope is as small as $M < 1.0M_{\odot}$.
For example, supernova 1993J in M81 has been identified as a type $\rm \uppercase\expandafter{\romannumeral2}$b \citep{Schmidt1993}. This behaviour implies that the progenitor of the core-collapse has a very small hydrogen mass at the time of
explosion, $\rm M_{H}\simeq 0.03-0.5M_{\odot}$ \citep[e.g.,][]{Woosley1994,Meynet2015,Yoon2017}, with a possible
mass down to $\rm M_{H} \simeq 0.001M_{\odot}$ \citep{Dessart2011,Eggleton1983}. In the population synthesis
investigation, \cite{Sravan2018} considered the mass of the hydrogen-rich envelope
of the progenitor at the onset of explosion to be
$0.01M_{\odot} \leq \rm M_{H} \leq 1.0 M_{\odot} $.

However, the mechanisms driving the stripping of the hydrogen envelope
and the parameter regimes that dominate the formation of SN $\rm \uppercase\expandafter{\romannumeral2}$b are still open questions.
Two physical mechanisms for envelope removal have been proposed to explain the progenitors of SN $\rm \uppercase\expandafter{\romannumeral2}$b. In the first scenario, a very massive star with initial mass ($\rm >25 M_{\odot}$)
is required in order for the mass-loss rate to be large enough \citep[e.g.,][]{Heger2003,Woosley1993} and at sufficiently high initial metallicity for stellar winds to be triggered. This scenario is supported by the analysis of the environments of type $\rm \uppercase\expandafter{\romannumeral1}$b/c
SNe, in which very massive stars
($\rm \geq 30 M_{\odot}$) have lost their envelopes through stellar winds \citep{Maund2018}. In order to meet with the complete set of observations SN 2008ax, \cite{Georgy2012} have constructed a type $\rm \uppercase\expandafter{\romannumeral2}$b progenitor of $20 M_{\odot}$ ending with the suitable core mass, color, luminosity, and hydrogen content. \cite{Groh2013} displayed the final stage of the rotating model as a luminous blue variable (LBV) star and proposed that LBVs may be the progenitors of some core collapse SNe. However, the observed stripping envelope SN rates are too high to be explained solely by single star evolution. The biggest difficulty with the single-star scenario is that the evolution of single star requires extremely precise fine tuning of the initial parameters to leave a thin hydrogen envelope prior to the explosion. Moreover, the clumping in stellar winds suggests that the currently used mass-loss rates are too high  {by single star evolution}; the hot wind mass-loss rates are lower by a factor of 2 or 3 than those typically used in stellar evolution calculations. The lower wind loss rate makes it difficult to produce SN $\rm \uppercase\expandafter{\romannumeral2}$b  {by single star evolution}.

The second scenario is close binary interactions involving
mass transfer via the Roche lobe overflow (RLOF) and  {possibly common-envelope evolution\citep[e.g.,][]{Podsiadlowski1992, Podsiadlowski1993, Nomoto1993, Yoon2010, Yoon2017}}, stellar evolution with rotation \citep[e.g.,][]{Georgy2012,Groh2013}, and nuclear burning
instabilities \citep[e.g.,][]{Arnett2011,Strotjohann2015}.
The typical ejecta mass of the stripped-envelope SNe may be very low ($\sim 2-4 M_{\odot}$), indicating that
the progenitors originated from lower-mass stars ($ \leq 20 M_{\odot}$) have lost their envelopes through binary interaction \citep{Lyman2016, Taddia2018, Prentice2019}.  {\cite{Joss1988} has shown that binary evolution can generate stripped supergiants with small H-rich envelopes. \cite{Claeys2011} have identified binary progenitor models for extended type $\rm \uppercase\expandafter{\romannumeral2}$b SNe. The hydrogen envelope of the primary star is stripped by the RLOF and only several tenths of solar mass of hydrogen envelope can be remained at the time of explosion.}

The most direct way to distinguish between these scenarios
is to search for a surviving binary companion after the supernovae. Such
searches have been successful for type $\rm \uppercase\expandafter{\romannumeral2}$b SNe, where putative surviving companions were discovered, e.g. SN1993J \citep{Maund2004}, SN2011dh \citep{Folatelli2014,Maund2019}, SN2001ig \citep{Ryder2018}. These findings are very important indicators which are originated from binary systems. \cite{Torrey2019} presented an enhanced mass loss scenario due to jets that the companion star might drive, and noted that an enhanced mass loss can cause the binary system to go through the grazing envelope evolution (GEE) and generate a progenitor of type $\rm \uppercase\expandafter{\romannumeral2}$b SNe. They estimated that the binary evolution channel with GEE contributes about a quarter of all SNe IIb. The GEE channel is completely different from the RLOF scenario, and hence widens the binary parameter space that can account for type $\rm \uppercase\expandafter{\romannumeral2}$b SNe. The fatal common envelope evolution scenario can also produce some $\rm \uppercase\expandafter{\romannumeral2}$b SNe. A less mass main sequence companion star circles inside the giant envelope of the primary star and eliminates most of the giant envelope before it merges with the giant core. However, this channel is confronted with some uncertainties in the calculations \citep{Soker2017,Lohev2019}.

Close binary stars are important in understanding the formation, evolution and death of massive stars. A high fraction of O-type stars ($70 \%$) at solar metallicity are expected to undergo a mass transfer episode during their lifetime \citep{Sana2012}. There is some evidence for two subclasses of SN $\rm \uppercase\expandafter{\romannumeral2}$b, those with radially-extended hydrogen envelopes and those with compact envelopes \citep{Chevalier2010}. \cite{Yoon2017} investigated the formation of SNe $\rm \uppercase\expandafter{\romannumeral2}$b in the channel of mass transfer via RLOF while considering three groups of
SNe $\rm \uppercase\expandafter{\romannumeral2}$b, namely, blue progenitors, yellow supergiants, and
red supergiants. The more compact blue progenitors and
yellow supergiants have hydrogen envelope masses less than about $\rm M_{H}<0.15M_{\odot}$, mostly resulting
from early Case B mass transfer with relatively low initial masses and/or low metallicity.
Red supergiants have hydrogen masses of $\rm M_{H }>0.15M_{\odot}$ at
the explosion, which can be produced via late Case B mass transfers.


In this paper, we intend to explore how the close binary evolution scenario can produce in producing the SNe $\rm \uppercase\expandafter{\romannumeral2}$b.
We aim to explore the following questions in the binary scenario: 1) how some initial physical parameters (i.e., rotational velocities, overshooting, metallicity, and orbital period) impact on the formation of SNe $\rm \uppercase\expandafter{\romannumeral2}$b; 2) how the surface chemical abundance varies with these initial parameters; 3) what controls the mass of the H-rich envelope of the progenitor through mass transfer due to RLOF;  4) what is the relation between SN $\rm \uppercase\expandafter{\romannumeral2}$b and other types of supernovae, such as SNe $\rm \uppercase\expandafter{\romannumeral2}$P, $\rm \uppercase\expandafter{\romannumeral2}$L and $\rm \uppercase\expandafter{\romannumeral1}$b/c; 5) how the internal structure of the deep core is influenced by these initial parameters.

In Section 2, we describe the physical ingredients of the stellar models and the domain of initial conditions that we have explored in this work.
In Section 3, the results of numerical calculations for the evolution of single stars and binary systems are presented in detail.  {In Section 4, we discuss an unsolved, long-standing problem about the discrepancy between the observed ratio of type $\rm \uppercase\expandafter{\romannumeral2}$b SNe and the theoretical one.}
Conclusions and summaries are given in Section 5.
\section{The initial parameters and model descriptions}

All models are calculated
with the MESA code \citep{Paxton2011,Paxton2013,Paxton2015,Paxton2018}.
We make use of the Schwarzschild criterion to determine the boundaries
of the convective region. The mixing
length parameter is $\rm l_{m}=1.5 H_{\rm P}$, where $\rm H_{\rm P}$ is the pressure scale height at the outer boundary of the core.  {We consider
a overshooting parameter of 0.12  {$\rm H_{\rm P}$} as the reference value. \cite{Sravan2020} have shown that SN2013df and SN1993J have a very low mass limit (about 2.0-2.8 $M_{\odot}$) of helium cores. This fact makes a smaller overshooting parameter to be a more appropriate choice and a smaller overshooting parameter can also be supported from intermediate-mass eclipsing binaries \citep{Stancliffe2015}. Normally, the standard value of the overshooting is 0.25 of $\rm H_{\rm P}$. \cite{Brott2011} considered convective overshooting using a parameter of 0.335  {$\rm H_{\rm P}$}. This value results from their new calibration using the observed $\rm v \sin i$ drop that is found in their data when they plot $\rm v \sin i$ against the surface gravity.} We assume that the helium abundance increases linearly from
$Y = 0.2477$ \citep{Peimbert2007} at $Z = 0.0$ to $Y
= 0.28$ at $Z = 0.02$ \citep{Brott2011}. We adopt the
basic.net, coburn.net, and approx21.net nuclear
networks in MESA.

Our models are comprised of single or two zero-age main sequence (ZAMS) stars and various initial parameters are listed in Table 1.
 {In order to conform with observations of two component stars of SN 1993J in the HR diagram, the accretion efficiency $\rm \beta_{mt}$, (i.e. the fraction of transferred
material that is accreted by the companion star) is then chosen as $\rm \beta_{mt}=0.5$. The final hydrogen envelope mass increases if we employ a lower accretion efficiency. For a higher accretion efficiency, the secondary star tends to evolve towards an over-luminous O star.} The non-accreted matter is directly expelled out
from the system as a fast wind for the accretor and carries the
specific orbital angular momentum of the mass gainer. We use the Dutch scheme in MESA for
both hot and cool wind mass loss rates, with the Dutch scaling
factor of 1.0\footnote{The Dutch wind mass-loss scheme is a combination of the prescriptions
of \cite{Vink2001} (when $\rm T_{eff} \geq 10^{4} $ K and $\rm X_{surf} \geq
0.4$ ), \cite{Nugis2000} (when $\rm T_{eff} \geq 10^{4} $ K and $\rm X_{surf} <
0.4$ ), and \cite{de1988} (when $\rm T_{eff} < 10^{4} $ K)}. The wind of Wolf Rayet stars is computed according to \cite{Nugis2000}. Radiative opacities were interpolated from the OPAL tables \citep{Iglesias1996}. The opacity increase due to Fe-group elements at $\rm T \sim 180 $ kK plays an
important role in determining the envelope structure in our stellar models.

We take into account various instabilities induced
by rotation that result in the mixing of chemical elements: Eddington-Sweet circulation,
dynamical and secular shear instability, and the Goldreich-Schubert-Fricke
instability  {\citep{Endal1978, Pinsonneault1989, Maeder2000a,Maede2012}}.
The rotational mixing owing to these hydrodynamic
instabilities is considered as diffusion processes
according to \cite{Heger2000}. The diffusion coefficients are adopted for the transportation of both chemical species
and angular momentum.
The contribution of the rotationally
induced instabilities to the total diffusion coefficient of the chemical species is
decreased by the parameter of $\rm f_{c}=0.0228$. This factor has been
calibrated to reproduce the observed nitrogen surface abundances as a function of the projected rotational velocities for
stars in the Large Magellanic Cloud sample (NGC 2004) of the
FLAMES survey \citep{Brott2011}.  {The parameter $\rm f_{\mu}$ denotes the sensitivity of the rotationally induced mixing to mean molecular weight gradients (i.e., the mean molecular weight gradients $\rm \nabla_{\mu}$ is replaced by $\rm f_{\mu}\nabla_{\mu}$) (Heger, Langer \& Woosley 2000).}
We adopt a value $\rm f_{\mu}=0.1$ as in \cite{Yoon2006} who calibrated this
parameter to match the observed surface helium abundances in stellar models with the solar metallicity.

The upper mass limit of the hydrogen envelope for SNe $\rm \uppercase\expandafter{\romannumeral2}$b heavily depends on the supernova
parameters, such as chemical composition, total mass of ejecta, and supernova energy. We assume that models with final H-rich envelope mass more than
$\rm 0.5 M_{\odot}$ explode as SNe $\rm \uppercase\expandafter{\romannumeral2}$P or SNe $\rm \uppercase\expandafter{\romannumeral2}$L, while those with envelope
mass less  {than $\rm 0.033 M_{\odot}$} explode as SNe Ib or Ic. Models whose
envelope mass is  {between $\rm 0.033 M_{\odot}$} and $\rm 0.5 M_{\odot}$ are adopted as SN
$\rm \uppercase\expandafter{\romannumeral2}$b progenitors. The properties of the SN $\rm \uppercase\expandafter{\romannumeral2}$b progenitor are also shown in Table \ref{table1}. The final evolutionary positions in the HR diagram are classified according
to their effective surface temperature and surface hydrogen mass fraction, as follows: Red supergiant (RSG): $\rm T_{\rm eff} < 4.8kK, X_{s}\geq 0.01$; Yellow supergiant (YSG): $\rm 4.8kK < T_{\rm eff} < 7.5kK, , X_{s}\geq 0.01$; Blue supergiant (BSG): $\rm 7.5 kK <T_{\rm eff} < 55 kK, , X_{s}\geq 0.01$. Hot helium giant (HeG): $\rm 15 kK < T_{\rm eff}  < 55 kK, X_{s}< 0.01$; Cool helium giant: $\rm T_{\rm eff} < 15 kK,  X_{s}< 0.01$; Wolf-Rayet star (WR): $\rm  10 kK < T_{\rm eff} < 251 kK, X_{s} \leq 0.4$ \citep{Gilkis2022}.

The initial parameters for single stars and the binary system are listed in Table \ref{table1}. The binary orbit is assumed to be circular and the Roche lobe radius is given by the formula of \cite{Eggleton1983}. The mass ratio is set to be $\rm q=0.882$ for all binary models.  {In the systems with $q=\frac{M2}{M2}< 0.7-0.8$, relying on the orbital period,
the mass-transfer rate via RLOF is so large that two component stars come into contact. Further common envelope evolution of these models requires complex considerations which are beyond the scope of this paper.
Actually, binary systems with lower mass ratios have problems to explain SNe $\rm \uppercase\expandafter{\romannumeral2}$b, in particular those with extended hydrogen envelopes having $M_{H}> 0.15 M_{\odot}$ \citep{Podsiadlowski1992}.}
We choose several initial orbital periods corresponding to cases where the first mass transfer event occurs during the main sequence phase ($\rm P_{orb}=3.0$ days, Case A), after core H-exhaustion but before the He-ignition in the core ($\rm P_{orb} =10.0$ days, Case B), during the core He-burning ($\rm P_{orb}> \sim 40.0$ days, Case C).
\begin{deluxetable}{lccccccccccccc}
\tablecaption{The parameters adopted in our calculations.
\label{table1}}
\tablewidth{0pt}
\tablehead{
\colhead{Models} & \colhead{$M_{\rm 1,ini}$}& \colhead{$M_{\rm 2,ini}$} & \colhead{$V_{\rm 1,ini}$} & \colhead{$V_{\rm 2,ini}$} & \colhead{$P_{\rm orb,ini}$}& \colhead{$\rm \alpha_{over}$} & \colhead{Z} & \colhead{$\rm M_{He}$} &\colhead{$\rm M_{H}$} & \colhead{$\rm R/R_{\odot}$}&\colhead{ST}&\colhead{SP}\\\hline
\colhead{} & \colhead{$M_{\odot}$} & \colhead{$M_{\odot}$} & \colhead{km/s} & \colhead{km/s} & \colhead{days} &\colhead{} &\colhead{} &  \colhead{$M_{\odot}$} & \colhead{$M_{\odot}$}  & \colhead{} &\colhead{} &\colhead{} }
\startdata
S1  &15   &..   &0   &..     &..            & 0.12& 0.02& 4.34& 8.60&741 &\uppercase\expandafter{\romannumeral2}P&RSG &\\
S2  &15   &..   &200 &..     &..           &0.12 & 0.02& 4.40& 7.62& 812&\uppercase\expandafter{\romannumeral2}P&RSG &\\
S3  &15   &..   &400 &..     &..            & 0.12 & 0.02&5.87 & 3.52& 1023&\uppercase\expandafter{\romannumeral2}P&RSG &\\
S4  &17   &..   &0   &..     &..            & 0.12& 0.02& 5.19&8.16 & 950&\uppercase\expandafter{\romannumeral2}P&RSG &\\
S5  &17   &..   &200 &..     &..            &0.12 & 0.02& 5.21& 8.02& 948&\uppercase\expandafter{\romannumeral2}P&RSG &\\
S6  &17   &..   &400 &..     &..            & 0.12 & 0.02&6.38& 3.67& 891&\uppercase\expandafter{\romannumeral2}P&RSG &\\
S7  &19   &..   &0   &..     &..            & 0.12& 0.02&6.00 & 9.00& 1047&\uppercase\expandafter{\romannumeral2}P&RSG &\\
S8  &19   &..   &200 &..     &..            &0.12 & 0.02&5.92 & 8.11& 1061&\uppercase\expandafter{\romannumeral2}P&RSG  &\\
S9  &19   &..   &400 &..     &..            & 0.12 & 0.02&7.30 & 3.13& 851&\uppercase\expandafter{\romannumeral2}P&RSG &\\
B1  &17  &15 &0   &0    &300.00        &0.12 &0.02 & 4.81& 0.35& 478&\uppercase\expandafter{\romannumeral2}b&RSG &\\
B2  &17  &15 &0   &0    &300.00        &0.25 &0.02 &5.61& 0.3& 407&\uppercase\expandafter{\romannumeral2}b&YSG &\\
B3  &17  &15 &0   &0    &300.00        &0.35 &0.02 &5.44& 0.0& 2&\uppercase\expandafter{\romannumeral1}b&WR &\\
B4  &17  &15 &0   &0    &300.00        &0.12&0.008 &5.54& 0.46& 512&\uppercase\expandafter{\romannumeral2}b&RSG&\\
B5  &17  &15 &0   &0    &300.00        &0.12 &0.03& 5.00& 0.27& 436&\uppercase\expandafter{\romannumeral2}b&YSG&\\
B6  &17  &15 &0   &0    &3.00        &0.12 &0.02 &3.34& 0.0& 6&\uppercase\expandafter{\romannumeral1}b&WR &\\
B7  &17  &15 &0   &0    &10.00        &0.12 &0.02 &4.60&0.14 & 14&\uppercase\expandafter{\romannumeral2}b&BSG &\\
B8  &17  &15 &0   &0    &700.00        &0.12 &0.02 &4.94& 0.49& 549&\uppercase\expandafter{\romannumeral2}b&RSG &\\
B9  &17  &15 &0   &0    &1600.00        &0.12 &0.02 &4.95& 1.88& 891&\uppercase\expandafter{\romannumeral2}P&RSG&\\
B10  &17  &15 &200   &200    &300.00        &0.12 &0.02 & 4.81& 0.35& 331&\uppercase\expandafter{\romannumeral2}b&YSG &\\
B11  &17  &15 &400   &400    &300.00        &0.12 &0.02 & 4.81& 0.35& 2&\uppercase\expandafter{\romannumeral1}b& WR &\\
B12  &17  &15 &0   &0    &50.00        &0.12 &0.02 & 4.63& 0.14& 166&\uppercase\expandafter{\romannumeral2}b& BSG &\\
B13  &16  &15 &0   &0    &1100.00        &0.12 &0.04 & 5.01& 0.39& 565&\uppercase\expandafter{\romannumeral2}b& RSG &\\
\enddata
\tablecomments{
\\
The meaning of each column is as follows.
  The symbol S denotes single stars whereas the symbol B denotes the binary systems.
  $M_{\rm 1,\rm ini}$: the initial mass of the primary star; $M_{\rm 2,\rm ini}$:
  the initial mass of the secondary star; $\rm V_{\rm 1,\rm ini}$: the initial equatorial
  velocity of the primary star; $V_{\rm 2, \rm ini}$: the initial equatorial velocity of the
  secondary star; $P_{\rm orb,\rm ini}$: the initial orbital period; $\rm \alpha_{over}$:
  the convective overshooting parameter; $\rm Z$: metallicity; $\rm M_{He}$: the mass of helium core
  at core carbon exhaustion; $\rm M_{H}$:the mass of hydrogen envelope at core carbon exhaustion. ST: Supernovae type. SP: the type of Supernovae progenitor.}
\end{deluxetable}
\setlength{\tabcolsep}{5mm}{
\begin{deluxetable}{ccc}
\tablecaption{The observations of SN 1993J and the theoretical values in model B1 and B13.
\label{table2}}
\tablewidth{0pt}
\tablehead{
 \colhead{Observations$^{1)}$} & \colhead{Model B1} & \colhead{Model B13} }
\startdata
$\rm \log T_{1, eff}=3.63 \pm 0.05$ &3.661 & 3.63\\
$\rm \log L_{1}/L_{\odot}=5.1 \pm 0.3$ & 4.968 &4.98\\
$\rm \log T_{2, eff}=4.3 \pm 0.1$ &4.54 &4.39\\
$\rm \log L_{2}/L_{\odot}=5.0 \pm 0.3$ & 4.87& 4.74\\
$\rm R/R_{\odot}\sim 600$ & 484&565\\
$\rm \dot{M} \sim 2-6 \times 10^{-6} M_{\odot}/yr$ &$\sim 2.58\times 10^{-6} M_{\odot}/yr$&  $\sim 3.0 \times 10^{-6} M_{\odot}/yr$\\
$\rm M_{H}=0.15-0.4 M_{\odot}$&$\rm M_{H}=0.35 M_{\odot}$ &$\rm M_{H}=0.39 M_{\odot}$\\
$\rm M_{He}=2.8-6.0 M_{\odot}$&$\rm M_{He}=4.81 M_{\odot}$ &$\rm M_{He}=5.01 M_{\odot}$\\
\enddata
\tablecomments{
\\
 ${}^{1)}$ The observational data of the progenitor SN 1993J is taken from \citep{Maund2004,Sravan2020}}
\end{deluxetable}}
\section{Results of numerical calculations}

We present non-rotating and rotating single star models and compare them with binary models with various initial parameters. We focus our investigation on the evolution of the primary star and explore whether close binary evolution via different initial orbital periods (i.e., Case A, Case B or Case C mass transfer), overshooting parameters, and metallicities could give rise to diverse Supernovae $\rm \uppercase\expandafter{\romannumeral2}$b in terms of the amount of the removed hydrogen or helium envelope. The evolution of the close binary system composed of a 17$M_{\odot}$ primary star and a 15$M_{\odot}$ companion star is computed. In all models, we calculate the evolution at least to the end of central Neon burning.

Properties of single stars and the primary star in binaries, such as evolutionary age, actual mass, radius, effective temperature, luminosity, central temperature and central density, ratio of the surface nitrogen to the initial value, equatorial velocities; mass fraction of chemical elements such as surface mass fraction of hydrogen and helium, logarithm of mass fraction of surface chemical elements such as carbon, nitrogen and oxygen, and the mass ratio of the surface nitrogen to carbon are presented in Table \ref{table3}.

\subsection{The evolution of the hydrogen envelope mass and surface nitrogen enrichments}

\subsubsection{Rotation effect}
Panel (a) in Fig. \ref{fig:general 1} shows the mass of the hydrogen envelope for single stars with $Z=0.02$ but with different initial masses and rotational velocities as a function
of the evolutionary age. It can be seen that the more massive star has a thicker hydrogen envelope at the end of the core H-burning phase. For example,
the mass of the hydrogen envelope for model S1 with $\rm 15.0 M_{\odot}$ is $\rm 11.38 M_{\odot}$  whereas it is $\rm 13.38 M_{\odot}$  for model S7 with initial mass of $\rm 19.0 M_{\odot}$.
Actually, stellar winds are stronger for more massive stars. The $\rm 19.0 M_{\odot}$ star  has lost more than 6 $M_{\odot}$ while the $\rm 15.0 M_{\odot}$ counterpart has lost less than 4 $M_{\odot}$.
In massive stars, mass loss via stellar winds is mainly a consequence of radiation
pressure on atoms during the main sequence and giant star stage.

The mass loss via stellar winds proportional to the luminosity of the star and is inversely proportional to
its effective temperature during central hydrogen burning. When massive stars evolve towards higher luminosities and
lower temperatures, a large fraction of the hydrogen envelope is actually lost by line driven winds
once the central hydrogen has been substantially converted in to
helium. The mass loss rates increase when the luminosity increases and hence when the initial mass increases. For a O type star, the total mass lost by winds during main sequence is approximately estimated by stellar mass at a power 2.8 (i.e., $\Delta M \approx M^{2.8}$). These single models undergo dramatic loss of mass
on the verge of core hydrogen exhaustion. This is due to the fact that higher luminosity triggers strong mass losses. Moreover, when some stars crosse some limit in effective temperature, there is an importance change in the ionization structure of the stellar envelope that may produce a great boost of the mass loss rate due to the bi-stability \citep{Vink2001}.

One can also note that most of the mass is lost during the
red supergiant phase of evolution when
the single star burns helium in its core.  After the core helium is exhausted, the mass of the envelope changes very little. This is because the evolution proceeds too fast to give rise to a significant mass loss.


Comparing model S4 with $\rm v_{ini}=0$ km/s and model S6 with $\rm v_{ini}=400$ km/s, one can
note that the mass loss is higher for rapidly rotating
stars during the main sequence. There are three reasons.
Firstly, mass loss via stellar winds can also be enhanced by the centrifugal force \citep{Langer1998}. The gravitational acceleration can be significantly reduced by the centrifugal force and
the star becomes more expanded. Rotation can also reduce the depth of the gravitational potential which stellar winds are easier to escape from and therefore increase the possibility of the formation of type $\rm \uppercase\expandafter{\romannumeral2}$b supernovae at the expense of the red supergiant stars. However, note that line driven stellar winds are powered by the radiative flux. This radiative flux is proportional to the effective gravity that decreases when the star is a rapid rotator. This behaviour favors the formation of polar winds. But this effects become much important only at velocities near the critical limit and are likely not important for models computed here. Secondly, rotation increases the main sequence lifetime and thus can allow more time for mass to be lost by stellar winds. Thirdly, rotating stars are more luminous than non-rotating or slower rotating ones and thus they undergo more mass loss by stellar winds.  Actually, the rotational mixing is the determining factor here. It changes the track in the HR diagram and makes the star to follow a different mass loss history.
Therefore, the rotationally enhanced stellar winds can decrease the minimum mass required for a single star to remove its hydrogen envelope
\citep{Meynet2003},
thus increasing the generation rate of SN $\rm \uppercase\expandafter{\romannumeral2}$b
progenitors from single massive stars.

At the end of the evolution, the mass of the hydrogen envelope is thinner in the model with an initial higher velocity. This can be explained by three reasons. Firstly, the convective core can be enlarged greatly by the rotational mixing and thus result in a thinner envelope (cf. Fig. \ref{fig:general 3}). Secondly, rotating stars are more luminous (also in the RSG phase). This enhances the stellar winds during that phase. As a result the star may evolve away from the red supergiant region in the HRD and becomes a yellow or even a blue supergiant. Finally, the angular momentum transport efficiency is maximum in the convective region
due to the largest convective diffusion coefficient.

Figure \ref{fig:general 2}(a) displays the surface mass fraction ratio of nitrogen to carbon for the
single star as a function of evolutionary age. There is no surface nitrogen enrichment in the non-rotational models S1, S4, and S7 until the first dredge up appears.
The outer convective region can span the mass coordinate from 16.5 $M_{\odot}$ to 4.6 $M_{\odot}$ during the first dredge up and may move toward the position of the hydrogen-burning shell.
After that, the outer convective envelope shrinks rapidly and develops again from the mass coordinate 13.45 $M_{\odot}$ to 5.39 $M_{\odot}$. It may approach the mass position of the hydrogen-burning shell ($ \sim 5.37 M_{\odot}$). The CNO products in the hydrogen-burning shell are mixed by the convective dilution. As a result, the surface ratio of nitrogen to carbon increases during the red supergiant stage.

The nitrogen enrichment for S1, S4, and S7 can also be ascribed to the mass removal
of hydrogen envelopes via stellar winds after the main sequence.
\cite{Markova2018} noted that the envelope is really
stripped in the most luminous supergiants or red supergiants by the strong winds
($\rm \log L/L_{\odot} \geq 5.8 $ and $ \log \dot{M} [M_{\odot}/yr]\geq -5.4$ ). Mass loss may reveal the matter with the enriched nitrogen as the surface matter of
the star is peeled. For example, the stellar mass in model S4 reduces from 16.51 $M_{\odot}$ at the end of hydrogen core burning to 13.45 $M_{\odot}$ at the end of helium core burning due to the strong RSG stellar winds.
A higher ratio of nitrogen to carbon can appear at the surface of the more massive stars. This indicates that these two processes are more efficient in more massive star.

In comparison with the non-rotating counterpart, a significantly higher surface ratio of nitrogen to carbon can be produced by a higher degree of rotational mixing during the main sequence \citep[e.g.,][]{Meynet2000,Maeder2014,Chieffi2013,Limongi2018,Song2018}. The main effect of rotational mixing is to smooth the internal chemical gradients and to facilitate a more progressive arrival of internal nuclear products at the surface \citep{Georgy2012,Ekstrom2008}. In MESA massive stars, Eddington-Sweet circulation dominates other rotation induced instabilities during the
main sequence and makes the whole star maintain rigid
rotation. In the subsequent evolution, dynamical shear dominates other instabilities in the stellar interior.
\cite{Maeder2009} presented results suggesting that the behavior of the surface excess of nitrogen is a multivariate function (i.e., stellar mass, evolutionary age, projected rotational velocity, metallicity) for a single rotating star. As expected, nitrogen enrichment increases with the increasing of the initial rotational velocity, initial mass and evolutionary age during the main sequence due to a higher velocity of the meridional circulation (cf, Panel a in Fig. \ref{fig:general 2}). Therefore, rapid rotation can help us explain the nitrogen-rich circumstellar material with a ratio of $\rm N/C \approx 12.4$ in SN 1993J.
Nitrogen enrichments can be aided by two extra factors. First, strong stellar winds which are enhanced by rotation can remove the hydrogen envelope and expose the hydrogen-burning shell which is richer in nitrogen. Second,
rapid expansion during post main sequence results in larger differential rotation which can strengthen the shear instability. Thus the spin angular momentum transportation from the core to the envelope becomes more efficient, meaning that the outer layer can attain a higher rotational velocity which favors efficiently rotational mixing and mass loss by a higher luminosity. For instance, one can notice that the equatorial velocity of the model S6 can attain 144.64 $\rm km/s$ at the onset of central helium burning (cf, Table \ref{table2}).
More importantly, the nitrogen enrichment factor $\rm \frac{N}{N_{ini}}$ goes up from 8.49 to 10.02 in model S6 while it rises from 1.0 to 4.34 in model S4 during the first dredge-up. This implies that
rotational mixing might reduce the efficiency of the dredge-up due to the decrease in the opacity of the outer envelope. The lower opacity implies a smaller convective envelope, thus reducing the depth of convective dredge-up.

\subsubsection{Convective overshooting effect}

A larger convective overshooting parameter can cause the star to evolve redward further in the Hertzsprung-Russel diagram. The reason is that the overshooting can lead to a larger convective core and extend the main sequence to lower effective temperature and higher luminosity. Therefore, these effects can also give rise to a larger mass loss via stellar winds and a thinner hydrogen envelope after the zero age main sequence in panel (b) of Fig. \ref{fig:general 1}. For instance, at the end of the core H-burning phase,
the stellar mass of model B1 with $\rm \alpha_{over}=0.12$ is $\rm 16.46 M_{\odot}$ whereas it is $\rm 16.07 M_{\odot}$ for the model B3 with $\rm \alpha_{over}=0.35$.  {Note that the overshooting parameter of 0.35 of $\rm H_{p}$ is similar to what \cite{Brott2011} have deduced (i.e., $\rm \alpha_{over}=0.335$), actually.}

This implies that mass loss at this age is closely related to the increased luminosity due to the convective overshooting
but to a lesser extent, on the effective temperature. However, the evolutionary track of the star computed with overshooting is much more extended
toward lower effective temperatures at the core hydrogen exhaustion. This implies a higher mass loss in model B3 and thus
its mass is lower than the one of the model B1 (cf. Table \ref{table2}).

The hydrogen envelope decreases rapidly for a star with a larger overshooting parameter at the core hydrogen exhaustion. Actually, the mass loss via stellar winds strongly depends on the stellar luminosity when the effective temperature decreases below 22500K \citep{Vink2001}. For example, the
envelope mass for model B3 with $\rm \alpha_{over}=0.35$ is $\rm 10.66 M_{\odot}$  whereas it is $\rm 12.15 M_{\odot}$ with $\rm \alpha_{over}=0.12$. Larger core
makes the star evolve more rapidly to the red part of the HR
diagram after the MS phase. This means that a larger fraction
of the core helium burning phase occurs during the RSG phase
where strong mass loss occurs. These strong mass losses favor
a more rapid appearance of deep layers at the surface.
During the first episode of Roche lobe overflow (hereafter, RLOF), the hydrogen envelope of $\rm 11.08 M_{\odot}$ in model B1 is transferred to the companion star while the hydrogen envelope of  $\rm 9.36 M_{\odot}$  is transferred to the companion star in B3. The fact implies that the more mass the hydrogen envelope remains during the main sequence, the more matter can be removed by RLOF.

Figure \ref{fig:general 2}(b) shows the surface mass fraction ratio of nitrogen to carbon for the
primary star with the different overshooting parameters in the binary system as a function of the actual mass.
The surface ratio of $\rm ^{14}N/^{12}C$ in the binary system B1 can attain a higher value of 93.370 compared to the value of 2.305 in its single-star counterpart S4 at the core helium exhaustion (cf., Table \ref{table3}). This is mainly because the surface chemical composition can be changed by the mass removal via RLOF. After the first event of RLOF, the ratio of $\rm ^{14}N/^{12}C$ can attain a larger value of 112 in model B3 with a larger overshooting parameter. This is because the hydrogen burning shell is located above the larger helium core and thus is buried in a shallow position of the hydrogen envelope. The envelope can develop a smaller outer convective region in this model. Therefore, the duration of the hydrogen burning shell is very short because the hydrogen-burning shell can be easily exposed by RLOF. The results also show that severe stripping of the hydrogen envelope usually can give rise to a higher surface effective temperature.

\subsubsection{The effect of metallicity}
It is shown in the panel (c) of Fig. \ref{fig:general 1} that the decreasing of metallicity can trigger smaller mass loss via stellar winds
and produce less stripped progenitors of type $\rm \uppercase\expandafter{\romannumeral2}$b supernovae. The mass-loss via stellar winds scales as $\rm \dot{M}_{wind}\propto Z^{0.85}$ \citep{Vink2000,Vink2021}. Therefore, the total mass lost during the lifetime of the star is strongly correlated with the amount of metals present in the envelope.
The star with low metallicity evolves essentially at almost constant mass during most of the main-sequence phase because of the absence of the larger mass loss via stellar winds. For this reason, the star with a lower metallicity has a thicker hydrogen envelope in comparison with the counterpart with higher metallicity. For example, when the core hydrogen is exhausted, the mass of the hydrogen envelope of model B4 with $\rm Z=0.008$ is $\rm 13.09 M_{\odot}$  whereas it is $\rm 12.17 M_{\odot}$ for model B5 with $\rm Z=0.03$. During the first episode of RLOF, $\rm 11.43 M_{\odot}$ in model B4 with $\rm Z=0.008$ is transferred to the companion star while it is $\rm 9.81 M_{\odot}$ in the B3 with $\rm Z=0.03$. Therefore, we can infer that more mass can be transfer via RLOF for stars with the same mass by lower metallicity.

Figure \ref{fig:general 2}(c) shows the surface mass fraction ratio of nitrogen to carbon for the
primary star with different metallicities and rotational velocity as a function of the actual mass.
The surface nitrogen $\rm \log ^{14}N$ has the same value of -3.394 as the initial one in the model B4 with $Z= 0.008$ while it is -2.996 in model B1 with $Z=0.02$ before the onset of RLOF (cf., Table \ref{table3}).  At lower Z, there are less CNO elements, and thus the nitrogen abundance is anyway less (even the one resulting from CNO process). This implies that the surface nitrogen can not be suddenly enhanced by the first dredge-up and the bottom of the outer convection region does not touch the hydrogen burning shell. It is shown that surface $\rm ^{14}N$ in the binary system B4 with a lower metallicity can attain a higher value after RLOF compared to its counterpart B5 with higher metallicity. The reason is that the total amount of mass transfer via RLOF is higher in the binary system with low metallicity.

The abundance of nitrogen is proportional to the initial metallicity prior to the core helium burning.
The reaction of nitrogen via $\rm ^{14}N(\alpha, \gamma)^{18}F(e^{+} \nu) ^{18}O$ is an important
exoergic process during the central helium burning. Low metallicity has an important impact on the energy generation during
hydrogen shell burning via the CNO cycle, which also affects the
boundary condition for the helium core.

\subsubsection{The orbital period effect}
 In the panel (d) in Fig. \ref{fig:general 1}, It can be found that RLOF is very efficient to remove the mass of hydrogen envelope in contrast to stellar winds. Beyond the core hydrogen exhaustion, the mass loss via stellar winds is $\rm 4.3 M_{\odot}$ for the single star S4 whereas it is $\rm 11.08 M_{\odot}$ during the first episode of mass transfer in model B1. This indicates that mass loss via stellar winds in the lower massive star (i.e., $\rm M < 19 M_{\odot}$) is too weak to remove the thick hydrogen envelope and produces type $\rm \uppercase\expandafter{\romannumeral2}$b supernovae. The single star will explode as a type $\rm \uppercase\expandafter{\romannumeral2}$P supernova (cf., Table \ref{table1}).

The transferred mass of the hydrogen envelope is very sensitive to the initial orbital period of the binary system. We note that an initially tighter orbit can result in a
deeper stripping of the hydrogen envelope via RLOF. For example, the transferred matter is $\rm 9.23 M_{\odot}$ for model B6 with $\rm P_{orb}=3.0$ days whereas it is $\rm 6.34 M_{\odot}$  for model B9 with $\rm P_{orb}=1600.0$ days.
The B6 system with $\rm P_{orb}=3.0$ days undergoes strong Case A mass transfer, followed by a later episode
of Case B mass transfer, with the last bit of remaining hydrogen removed by
strong Wolf-Rayet winds during the later core-burning phases. This model can produce a low mass helium core with $\rm M=3.3 M_{\odot}$ at the
time of explosion, with a radius of about 4.26 $R_{\odot}$.
However, the model B6 cannot be consistent
with the observed hydrogen envelope mass of $0.03-0.5 M_{\odot}$ for type $\rm \uppercase\expandafter{\romannumeral2}$b because it completely loses any even thin hydrogen envelope.
It is also possible to produce fully stripped type $\rm \uppercase\expandafter{\romannumeral1}$b progenitors (cf., Table \ref{table1} and Fig. \ref{fig:general 6}) \citep{Yoon2015,Yoon2010}. The model B9 finally explodes as a $\rm \uppercase\expandafter{\romannumeral2}$P supernova because its hydrogen envelope remains sufficiently large to preserve the RSG structure. These facts implies that the relationship of SN $\rm \uppercase\expandafter{\romannumeral2}$b with other
types of supernovae, such as SNe $\rm \uppercase\expandafter{\romannumeral2}$P, $\rm \uppercase\expandafter{\romannumeral2}$L and Ib/c, is closely bound up to
the hydrogen envelope mass. Moreover, various types of $\rm \uppercase\expandafter{\romannumeral2}$b supernovae are also related to the hydrogen envelope mass.

The system with the region 300 days $\rm <P_{orb}<$ 700 days can give rise to RSG type SN $\rm \uppercase\expandafter{\romannumeral2}$b progenitor. The
initial period of 300 days roughly separates between the RSG type SN $\rm \uppercase\expandafter{\romannumeral2}$b progenitor and the type YSG  SN $\rm \uppercase\expandafter{\romannumeral2}$b progenitor.
The binary system B7 with an initial $\rm P_{orb}=30$ days can produce the BSG type SN $\rm \uppercase\expandafter{\romannumeral2}$b progenitor (cf., Fig. \ref{fig:general 6}).
Thus binary models with orbital periods range $\rm  \sim10$ days $\rm < P_{orb}<$ 700 days may turn into SNe $\rm \uppercase\expandafter{\romannumeral2}$b (cf., panel d in Fig. \ref{fig:general 5}).
Most importantly, the number of mass transfer via RLOF is closely related not only to the initial orbital period but also to the thickness of the hydrogen or helium envelopes. Beyond the core hydrogen exhaustion, a larger amount of the residual hydrogen in the envelope can expand to larger radius during the late evolutionary stages in contrast to a bare helium core. After the core helium burning, the envelope expansion becomes more significant for more compact stellar core.

Panel (d) in Fig. \ref{fig:general 2} shows the surface mass fraction ratio of nitrogen to carbon as a function of the orbital period for the
primary star in the binary system with the different initial orbital periods. One can see that the donor star in binaries
can experience the envelope peeling which can expose the inner layers of
CNO processed material to the surface, increasing the surface
abundance of nitrogen. An
initial tighter orbit results in more significant peeling of the
hydrogen envelope, and vice versa.
The system with the shortest orbital period has the highest surface
N/C ratio in our grid. The main reason is that
the hydrogen burning shell can be revealed early because much more hydrogen envelopes can be eliminated by
RLOF.

\subsection{The evolution of the convective core and helium core}
\subsubsection{The effect of rotation}
Panel (a) in Fig. \ref{fig:general 3} shows the convective cores of the non-rotating and rotating single stars as a function
of the evolutionary age.
The mass of convective core increases with the initial mass of the star because the size of the convective core is governed by
radiative pressure which is proportional to the quadrature of core temperature $\rm T^{4}$. Actually, at the onset of the evolution, the centrifugal force partially sustains the gravity but most of the equilibrium is due to pressure gradient. Therefore, the rotating star behaves like a less massive non-rotating one, and the mass of the convective core is smaller. This results in a lower luminosity in the HRD. The rotating star has a slightly denser and cooler core than the non-rotating one.

As the evolution proceeds, the rotation induced mixing starts to refuel the core with fresh hydrogen.
The mass of the convective core is larger in models with higher initial rotational velocity because the rotation induced mixing becomes very
efficient in rapidly rotating stars. Actually, two main physical processes are responsible for such a rotational mixing: meridional circulation and secular shear.
Meridional circulations which are scaled as the square of rotational angular velocity are mainly responsible for rotational mixing above the convective
core \citep{Maeder2000b, Song2018}. Rotational mixing can slow down the decrease in mass of the convective core and is similar to the behavior of overshooting.
The larger core induced by rotational mixing leads to a higher central temperature and a lower opacity in the outer envelope. Rotating stars have a larger convective core than the non-rotating ones.

Fig. \ref{fig:general 4} displays the mass of the helium core as a function of the evolutionary age. The mass of the helium core generally scales with the size of the hydrogen  convective core and goes up with the increasing of the initial rotational velocities and stellar masses (cf., panel a in Fig. \ref{fig:general 4}). Generally,
the larger the size of the convective core is, the larger the final helium core mass would be at the end of the main sequence. The
maximum size of the hydrogen convective core generally increases with the mass of the
star and the rotational velocities. Therefore, the helium core at core helium exhaustion can increase with the mass of the
star and rotational velocities as well.
Furthermore, the main consequence of the rotational mixing is the increase of the lifetime of the core hydrogen burning. The main reason is that fresh hydrogen in the outer envelope is transferred into the central core by rotational mixing. This mixing process increases the hydrogen fuel supply in the stellar core.

The wind mass-loss rate of single stars with $M \geq 30 M_{\odot}$ is strong enough to remove the hydrogen envelope and these stars are expected to generate the type $\rm \uppercase\expandafter{\romannumeral2}$b SNe \citep{Heger2003,Georgy2009}. This type of star has a helium core mass $\geq 8 M_{\odot}$  previous to the explosion. However, a $8 M_{\odot}$ helium core is too massive to produce the second maximum at $\sim 20$ days as observed light curves for SN 2011dh, even assuming the most extreme $\rm ^{56}Ni$ mixing \citep{Bersten2012}.
The helium core mass of single stars goes up because of the creation of helium by the hydrogen-burning shell (cf. panel a in Fig. \ref{fig:general 4}). However,
one can see that the mass of the helium core in model B6 can be greatly influenced by the RLOF during the main sequence (cf. panel d in Fig. \ref{fig:general 4}).
This is because the convective core can be greatly reduced by the RLOF. This fact indicates that for a given initial mass, the star in the binary system
has a smaller core mass than its single counterpart. However, the corresponding radii go up with the decreasing of the core helium mass \citep{Yoon2010,Yoon2017}.
After the main sequence, the evolution
of the helium core is almost unaffected by the mass transfer. The helium produced by the hydrogen burning shell contributes little to the mass of the core helium.
The binary-peeled star loses matter after the RLOF because of strong stellar winds, leading to a reduction in the helium core mass.
Therefore, the amount of oxygen that can be produced from a specific exploding core is smaller in an initial tighter system.

However, the mass fraction of the convective core in the slowly rotating massive stars with $\rm v_{ini}=200$ Km/s is slightly smaller than the non-rotating stars beyond the core hydrogen exhaustion. Rotational mixing plays a minor role in this case. During the core helium burning, the convective core grows largely because the
mass of the helium core itself grows regardless of the strong mass losses via the stellar winds. As the mass of the core grows, its luminosity increases accordingly but the radius of the convective core remains constant. Therefore, the mass-loss rate of stellar winds can grow rapidly. Meanwhile, the central temperature and density go up accordingly. This
favors the formation of convection.  The helium burning convective core
grows until close to the time of the core helium exhaustion because the central temperature increases due to the decrease of the fuel supply. This actually increases the energy generation rate and thus the convective core. This growth of the helium core can have a
very important consequence. The addition of helium to the helium convection
zone at late time increases the O/C ratio made
by helium burning.

\subsubsection{Convective overshooting  effect}
The overshooting can cause its convective core to grow in
mass and this leads to the mixing of fresh hydrogen above the core into the central nuclear-burning zone. Therefore, the overshooting has the effect of extending the main-sequence lifetime.
The band of main sequence extends to lower effective temperature values when convective core overshooting is
larger. A massive star with initial $\rm M > 19 M_{\odot}$ has a helium core $\rm M_{He}> 6 M_{\odot}$ prior to the SN explosion (cf., panel a in Fig. \ref{fig:general 4}).
This fact indicates that the low mass progenitor of SNe $\rm \uppercase\expandafter{\romannumeral1}$b only can be produced in the maximum mass limit $\sim 19 M_{\odot}$.

The amount of helium in the envelope can also be convected into the helium core near central helium exhaustion by the convective overshooting. At the stage of the advanced nuclear burning, the higher overshooting parameter leads to a sightly higher helium and carbon core masses (cf., Panel c in Fig. \ref{fig:general 4}). However, the lifetime during helium burning might be reduced by the convective overshooting because the central temperature can be increased. The efficiency of helium combustion becomes higher because of the enlarged helium core.
For example, the lifetime of the core helium burning in model B1 with $\rm \alpha_{over}=0.12$ is about 0.97 Myr whereas it is about 0.76 Myr in the model in model B3 with $\rm \alpha_{over}=0.35$ (cf., Table \ref{table3}).

\subsubsection{The effect of metallicity}
The effect of metallicity on the convective core mass is illustrated in Panel (c) in Fig. \ref{fig:general 3}. At the beginning of the evolution, the convective core decreases with the decreasing of the metallicity.
The reason is that a lower
initial metallicity also indicates a reduction of the abundance of the CNO nuclei and
therefore a decrease of the hydrogen burning efficiency. However, it is shown in the second half of the main sequence and the situation has been changed.
In order to maintain higher luminosity, the star has to contract more to increase the core temperature and the nuclear energy production. The convective core can be enlarged by the increased central temperature.
Moreover, the initial metallicity has an important impact on initial hydrogen and helium abundances. For the lower metallicity, there is more hydrogen to
burn in the central core. The lifetime of the main sequence can be enlarged by the fresh fuel supply.

The effect of metallicity on the helium core mass is shown in Panel (c) in Fig. \ref{fig:general 4}.
\cite{Limongi2018} presented that the less massive star (i.e., $\rm < 40 M_{\odot}$) develops helium core masses essentially independent of the initial metallicity in single stars.
This implies that the mass of the helium convective core weakly depends on the metallicity.
However, we can see that a smaller mass of the helium convective core (or helium core) can come into being in a low metallicity environment after core hydrogen depletion (cf., Panel C in Fig. \ref{fig:general 3}).
Moreover, in massive stars with low metallicity, the
helium convective core can grow so much during the central helium burning that
it approaches the hydrogen burning shell.
Furthermore, the helium convective core never recedes until the core helium exhaustion. The reason is that the conversion of helium to carbon
and oxygen increases the opacity in the entire convective core
so that its outer border is continuously pushed outward
rather than inward. Hence, a strong chemical discontinuity forms at the border of the helium convective core where
the helium abundance changes from roughly zero to roughly
one at central He exhaustion. A helium convective
shell starts to form outside the maximum extension of the
convective core whereas the chemical composition of this region
is still the one left by hydrogen burning.

\subsubsection{The effect of the orbital period}
From panel (d) in Fig. \ref{fig:general 3},  can also see notice that the mass of the convective core drops from 4.44 $M_{\odot}$ to 2.73 $M_{\odot}$ during the first episode of RLOF for model B6 with initial $\rm P_{orb}=3.0$ days. The result shows that the primary star that loses its hydrogen envelops via RLOF will develop a smaller convective core than its single counterpart. This fact indicates that RLOF can accelerate the mass decrease of the convective core during main sequence.
The reason is that the core temperature can be reduced significantly by RLOF. However, mass transfer via the RLOF has the effect of extending the main-sequence lifetime of the primary star because the efficiency of hydrogen burning via CNO cycle is extremely sensitive to the core temperature.
The convective core after main sequence is also influenced by the previous RLOF because it displays a smaller value after main-sequence than the counterpart with initial $\rm P_{orb}=300$ days in model B1.

The effect of the orbital period on the helium core mass is shown in Panel (c) in Fig. \ref{fig:general 4}.
The helium core formed in binary systems is less massive than the single counterpart by comparing the model S4 with B6. This can be explained by an early onset of
Case A mass transfer in this system and a significant loss of
matter during this process. Case A mass transfer takes
place in the primary star of model B6 before the helium core is fully formed.
Mass loss via RLOF is strong enough to decrease substantially
the total mass and therefore to induce a reduction of the convective core during
the main sequence.
The primary star in this system loses about
12.8 $M_{\odot}$ during all binary interactions via RLOF.
As a consequence, in this case, the helium core at core He
depletion is smaller than it would be in Case B or Case C mass transfer.

Actually, the evolution of the star after core hydrogen burning mainly depends on the mass of the helium core rather than
the total mass. This property
implies that mass loss via RLOF globally affects the evolution during the main sequence because it is efficient enough
to determine the reduction of the mass of the hydrogen
convective core, which in turn is mainly
responsible for the initial mass of the helium core.
A small convective core develops in an initial tighter binary system.
RLOF in Case B or Case C has a little impact on the evolution of the convective core (cf., Panel d in the Fig. \ref{fig:general 4}).
The growth of the mass of helium cores can be merely affected by the development of the hydrogen burning shell. However,
the shell can be removed or extinguished by the RLOF, as shown in model B6 with an initial $\rm P_{orb}=3.0$ days.

\subsection{The evolution of the rate of mass transfer via RLOF}

\subsubsection{The effect of convective overshooting}

When the primary star expands beyond its Roche lobe, mass is transferred to the companion.  {It is the expansion of the donor due to its own nuclear evolution that initiates mass transfer. Actually, the expansion of stellar radii makes the primary star continue to fill its Roche lobe and to trigger mass transfer via RLOF.}
From panel (a) in Fig. \ref{fig:general 5}, can see that the maximum rate of mass transfer is closely related to the convective overshooting. The maximum rate of the mass transfer via RLOF increases with the overshooting parameter. This can be understood by the fact that the larger radii can be induced by the larger convective overshooting parameter and the mass transfer rate has an exponential dependence on stellar radius. The first maximum value of the mass transfer rate heavily depends on the termination of decreasing of the orbital period at mass ratio one $\rm q=\frac{M_{2}}{M_{1}}$. When the primary star has a radiative envelope in Case A and early Case B mass transfer, the mass transfer via RLOF happens on the Kelvin-Helmhotz timescale. The primary shrinks rapidly in response to mass loss (i.e., the adiabatic mass-radius exponent $\rm \zeta_{ad}=(\frac{d\log R}{d\log M})_{ad}\gg 0$) while the Roche lobe
contracts in response to mass loss as well (cf., Panel f in Fig. \ref{fig:general 6}). RLOF can proceed only if the Roche lobe is slightly smaller than the stellar radius.
One can find that the number episode of mass transfer via RLOF might decrease with the increasing of the overshoot parameter. It is closely related not only to the residual amount of the hydrogen envelope but also to the orbital period of the system. Moreover, the expansion of the helium envelope beyond the core carbon
exhaustion becomes more prominent for a more compact carbon-oxygen core via the mirror effect.

In fact, the final orbital period of the binary system goes up with the total amount of mass transfer via RLOF. The RLOF ceases when the donor lost
most (but not all) of its hydrogen rich layers. The primary star still has some hydrogen
left in its envelope (typically a few $\rm 0.1 M_{\odot}$).
Moreover, the mass transfer occurs early
in the star with a smaller convective parameter. For example, the model B1 with $\rm \alpha_{over}=0.12$ occurs RLOF at $9.95$ Myr whereas the model B3 with $\rm \alpha_{over}=0.35$ occurs RLOF at $11.15$ Myr. The main reason is that the lifetime has been enlarged by the convective overshooting parameter.
The model with a higher overshooting is redder
and more extended than the one with a smaller overshooting and therefore is more prone to
go through stronger mass transfer episode during the RSG
phase.

Mass-transfer rates via RLOF can be very high (i.e., $\rm 10^{-2}-10^{-3} M_{\odot}/yr$),
and can far exceed any mass-loss rate for a line-driven wind. The binary evolutionary channel supports the formation of type $\rm \uppercase\expandafter{\romannumeral2}$b supernovae.
The observation of the binary companion in the case of the type $\rm \uppercase\expandafter{\romannumeral2}$b SN 1993J, and possiblely also in the case of SN 2013df, implies the binary channel. Although the RLOF is the main physical process for mass removal, it can not completely eliminate the whole hydrogen envelope from the primary star. Whether the SN progenitor can remain a large amount of hydrogen is closely related to the subsequent mass loss via stellar winds.
Stellar winds are stronger for larger overshooting and favor the formation the SN $\rm \uppercase\expandafter{\romannumeral2}$b (cf., panel b in Fig. \ref{fig:general 7}).



\subsubsection{Metallicity  effect}
From panel (b) in Fig. \ref{fig:general 5}, one can see that the maximum rate of mass transfer via RLOF also depends heavily on the metallicity.  The smaller the metallicity, the greater the maximum mass transfer rates. Actually, stellar winds for two components of the binary system tend to widen the
binary separation and reduces the total amount of mass lost by RLOF.
Stars with low metallicity have weaker stellar winds and more hydrogen envelopes available for mass transfer via RLOF. Thus, the size of the Roch lobe becomes smaller in the system with low metallicities.
More hydrogen can be retained before the onset of RLOF in the model with the lower metallicity. Therefore, less hydrogen is retained after the RLOF phase in the lower-metallicity model (cf., Panel c in Fig \ref{fig:general 1}). The mass transfer rate depends very sensitively on the fractional radius excess of the donor, $\frac{\Delta R}{R_{L}}=\frac{R_{D}-R_{L}}{R_{L}}$. $\rm R_{D}$ is the radius of the donor while $\rm R_{L}$ is the radius of the Roche lobe. The quantity $\rm \triangle R$ is the radius excess. This implies that the quantity $\rm \triangle R$ is larger for the star with lower metallicity.
One can also notice that the higher the metallicity, the earlier RLOF begins. For example, model B5 with $\rm Z=0.03$ begins RLOF at the age of 9.0 Myr whereas model B4 with $\rm Z=0.008$ begins RLOF at the age of 11.16 Myr. The main reason is that stars with higher metallicity have larger stellar radii.

It can be noticed that the star with low metallicity is prone to generate the more compact blue progenitors and remain less hydrogen mass at the end of the RLOF in contrast to the counterpart with high metallicity.
However, stellar winds after the RLOF are stronger for the star with high metallicity and it favors a higher effective temperature because the thin hydrogen envelop can be removed.

\subsubsection{The orbital period effect}

From panel (c) in Fig. \ref{fig:general 5}, one can see that the maximum rate of mass transfer via RLOF also depends heavily on the orbital period. One can notice that the mass transfer rate is larger for an initial wider binary system because the primary star is farther evolved.
Mass transfer becomes increasingly unstable and rapid from Case A mass transfer to Case C mass transfer. Case A mass transfer occurs in the system B6 with the shortest initial orbit period.
The first phase of mass transfer can usually be
divided into a rapid phase on the thermal timescale of the primary, followed by a slow phase on the much
longer nuclear timescale. The fast phase
of mass transfer proceeds until the primary regains thermal equilibrium, i.e.
when its equilibrium radius becomes smaller than its Roche radius.

For Case A mass transfer, the orbit widens and the mass transfer rate falls down as the mass ratio of the primary to the secondary $\rm q=\frac{M_{2}}{M_{1}}$ approaches unity. Beyond this time the donor star
has become the less massive star in the system, in other words, the mass ratio has been reversed. During the rapid mass-transfer phase, both stars are out of thermal equilibrium: the
primary is somewhat less luminous due to mass removal, while the secondary is somewhat more luminous
as a result of accretion.
The mass transfer rate is also sensitively dependent on the mass ratio of the secondary star to the primary star $q=\frac{M_{2}}{M_{1}}<1$, with increasing rates for lower values of mass ratio q.

The second RLOF occurs
at 10.635 Myr when the envelope of the primary star
expands due to hydrogen shell burning during the helium core
contraction phase.
The resulting high-mass transfer rate (i.e., $\dot{M}_{R} \sim 1.9953\times 10^{-4} M_{\odot}/yr$ ) occurs again on the Kelvin-Helmholtz timescale. The primary star
loses most of its hydrogen envelope, exposing its
helium core of $3.5 M_{\odot}$ with a small amount of hydrogen envelope of $\rm M_{H} = 1.1 M_{\odot}$. The luminosity $\rm \log L/L_{\odot}$ increases from 4.52 to 4.72 accordingly.
During this phase of mass transfer, the orbital
separation increases significantly.
Although the star remains compact ($R \sim 1.56 R_{\odot}$) at the terminal of the second mass transfer, the helium shell burning can be activated after core
helium exhaustion. This leads to the expansion of the envelope up
to $\sim 6.3 R_{\odot}$ during core carbon burning.

Case B mass transfer in massive binaries B7 with an orbital periods of $\rm \sim 10$ days is similar to Case A mass transfer in many respects. Because the
envelope of the primary star is radiative, the mass transfer starts with a rapid, thermal-timescale phase during which the
mass ratio is reversed. An important difference is that since the donor star is more extended and therefore
has a shorter thermal timescale compared to Case A mass transfer in model B6, and the mass transfer rates are correspondingly larger during this phase. Moreover, the primary star is itself in a rapid phase of evolution when
it passes through the Hertzsprung gap. It is out of thermal equilibrium and expands on the timescale at which its
core contracts. As a consequence, after mass-ratio reversal mass transfer continues on the expansion
timescale of the primary, only slightly slower than the thermal timescale. Therefore there is a lack of slow
phase of mass transfer as in Case A mass transfer. Mass transfer continues at a fairly high rate until most of the
envelope has been removed. The evolutionary track forms a loop in the HR
diagram during the mass transfer phase, and the maximum transfer rate coincides with the
minimum luminosity of the donor star (cf., Fig. \ref{fig:general 7}). The decrease of the luminosity during mass transfer is
caused by the large thermal disequilibrium of the primary star. A primary star with the radiative envelope
shrinks in response to mass loss and has to expand again to reach thermal equilibrium. This requires the
absorption of the thermal energy, so that the surface luminosity during thermal-timescale mass transfer
is extremely smaller than the nuclear luminosity provided by the H-burning shell.

When central helium is ignited in the core, the mass transfer via RLOF ceases. The primary star contracts and detaches
from its Roche lobe. This occurs when the primary is almost a bare helium core with $\rm M_{He} \sim 4.15 M_{\odot}$ and a thin H-rich layer $\rm \sim 0.75 M_{\odot}$. The primary shifts to a position close to the helium main-sequence in the HR
diagram. The remnants of Case B mass transfer are in a long-lived
phase of evolution: these are binaries consisting of an almost bare helium-burning core primary
and a more massive main-sequence companion star. Mass transfer via RLOF will widen their orbits significantly.
Observed counterparts of this evolution phase among massive systems are the WR+O binaries,
consisting of a Wolf-Rayet star and a massive O star.


A common prescription for the largest rates of RLOF is to assume that the mass transfer is limited by the thermal timescale of a
massive star with a radiative envelope. This implies that the
mass-transfer rate is likely to be highest for more massive and more luminous stars, which have short thermal
timescales. The thermal time scale can also decrease as the star evolves. This indicates that the mass transfer rate via RLOF is higher for wider systems, in which
the donor star is more evolved.
Mass-loss rates can be of order $\rm 2.4 \times 10^{-2} M_{\odot} yr^{-1}$ for model B9 with an initial $\rm P_{orb} \approx 1600$ days. The main reason is that the adiabatic response of the
donor is unable to keep it within its Roche lobe, leading to ever-increasing mass-transfer rates \citep{Ritter1988,Wellstein2001}. Mass
transfer accelerates to a timescale in between the thermal and dynamical timescales of the donor. Stars
with deep convective envelopes , i.e. a giant or red supergiants, tend to expand or keep a roughly constant radius (i.e., the adiabatic mass-radius exponent $\rm \zeta_{ad}=(\frac{d\log R}{d\log M})_{ad} \leq 0$) when it loses mass adiabatically. This fact implies that the response to the mass transfer of a convective envelope is very different from that of a radiative envelope.

The Roche-lobe radius shrinks when the mass is transferred from a more massive to a less
massive star (cf., panel f in Fig. \ref{fig:general 6}). The response to mass loss via RLOF can cause the donor to overfill its Roche lobe by an ever larger amount and cause runaway mass transfer. The maximum mass transfer rate for Case C is the highest among three Cases of mass transfer. However, as the orbital period increases, the total amount of mass transfer via RLOF decreases. This indicates that the duration of RLOF becomes shorter in the system with an initial longer orbital period.
When the envelope of the primary star
expands, The second event of RLOF happens
at 10.519 Myr. The radius cannot expand because it is restricted by the Roche lobe.
Finally, the primary has a helium core mass with $\rm M_{He} \sim 4.95M_{\odot}$  and a thick H-rich layer of $\rm \sim 1.87 M_{\odot}$.
Therefore, this kind of mass transfer can not give rise to type  $\rm \uppercase\expandafter{\romannumeral2}$b due to the existence of a thick envelope.

\subsubsection{Rotation effect}
From panel (d) in Fig. \ref{fig:general 5}, one can see that the maximum rate of mass transfer via RLOF is also related to the rotation. One can see that rapid rotation can favor the maximum mass transfer rate. However, the total amount of mass transfer via RLOF reduces with the increasing of rotational velocities. Rotation can delay the beginning of Case B mass transfer. There are three main reasons. Firstly, rotational mixing can prolong the lifetime of the main sequence. Second, the orbital separation can be widened by the angular momentum transformation because more spin angular momentum can be transferred to the orbit. Third, rotation enhanced stellar winds can also widen the orbital separation.

\subsection{Stellar radii and the evolution in the HR diagram}

\subsubsection{Rotation effect}
Panel (a) in Fig. \ref{fig:general 6} shows the photospheric radius for the non-rotating and rotating single stars as a function
of evolutionary age. One can see that from the beginning of evolution, an increase in rotational speed leads to a decrease in radius. It can be understood that the star receives a small gravitational acceleration because of the sustaining effect of the centrifugal force and behaves like a non-rotating star with a slightly lower mass. The central temperature is smaller and the corresponding central density is larger. This results in a decrease in both luminosity and radius. Massive stars expand more as they cross the Hertzsprung gap and ignite helium as a red supergiant
while their radii swell slightly. The rotation effects are almost unnoticeable in the evolution of the radius after the main sequence because of the rapid expansion of the hydrogen envelope.

However, the rapidly rotating star S9 can experience very strong mass
loss via stellar winds and its hydrogen-rich envelope can be removed when the core helium is ignited. The radius decreases and
the star slightly shifts toward a Wolf-Rayet star. The massive single star ($\rm \sim 19 M_{\odot}$) with an initial velocity of 400 Km/s can approach the inferred luminosity and effective temperature of the type $\rm \uppercase\expandafter{\romannumeral2}$b progenitor star. This star finally explodes as a type $\rm \uppercase\expandafter{\romannumeral2}$-P supernova.

Panel (a) in Fig. \ref{fig:general 7} shows the evolutionary tracks in the HR diagram for the non-rotating and rotating single stars.
The effective temperature behaves as $\rm T_{eff}\propto M^{0.5-0.6}$ at the onset of the evolution. Therefore, the effective temperature increases with the stellar mass.
For the star with a moderate initial rotational velocity
of $\rm v_{ini} < 200 km/s$, the increase of the stellar wind with respect to the non-rotating case
is limited. Rotational mixing merely plays a minor role in decreasing the opacity of the envelope.
These models can maintain the star in the red supergiant state throughout the whole core helium burning phase.
However, an efficient rotational mixing in model S9 can increase the size of the helium core significantly and the post-main sequence luminosity of the fast
rotating star is higher, by approximate a factor of $\sim 2.5$ than that of a non-rotating
counterpart S7. Therefore, the post-MS luminosity of the rapidly
rotating star is higher than a non-rotating
counterpart. The enhanced mass loss by rotation in model S3 implies that tracks for the stars
do not attain as far to the right of the HR diagram as they do for the non-rotating star S1. Furthermore, The star S9 with higher rotational velocities shifts its color earlier from the red part of HRD after the core helium exhaustion than the model S7 which does
not include rotation. The star terminates its evolution within the different regions of the HR diagram which have been defined by various types of SN $\rm \uppercase\expandafter{\romannumeral2}$b progenitors.
The rotating model S9 can approach the observed effective temperature of the type RSG SN $\rm \uppercase\expandafter{\romannumeral2}$b.
However, the luminosities of SN $\rm \uppercase\expandafter{\romannumeral2}$b are in the range of $\rm \log L/L_{\odot}=4.92-5.12$. The final luminosity of the model S9 is beyond the scope of the type $\rm \uppercase\expandafter{\romannumeral2}$b observations.

Therefore, less massive stars (i.e., $\rm M < 19 M_{\odot}$) with  higher rotational velocity are prone to form type $\rm \uppercase\expandafter{\romannumeral2}$b supernova. Furthermore, we also note that the blue-ward excursion can be increased by rotation in the binary system (cf. panel e in Fig. \ref{fig:general 7}).


\subsubsection{Overshooting effect}
One can see in Panel (b) of Fig. \ref{fig:general 6} that convective overshooting from the core evidently provides a larger
reservoir of available nuclear fuel during the main sequence and consequently it produces an increase of
the effective temperature and the luminosity after the RSG stage. Hence the radius of a star slightly decreases at
a given luminosity for a larger overshooting. During helium burning the star has higher overshooting parameter and
a smaller radius because a larger amount of the hydrogen envelope has been removed in model B3. At a smaller hydrogen envelope mass $\rm M_{H}< 0.1 M_{\odot}$, a higher luminosity always associates with a smaller radius in model B3. The radius remains smaller than it is at the tip of the giant branch. Before the core helium exhaustion, the radius of model B3 can display a significant expansion because of the existence of a thick helium envelope. A double-peaked light curve has been observed in SN 1993J. The light curve shows a rapid decline during one to three days after an initial peak, which is a consequence of cooling after the shock breaks out from the surface. The duration of the cooling phase is mainly dominated by the radius of the progenitor. More compact sizes of the progenitor, like Wolf-Rayet stars, give rise to faster declines than extended progenitors, such as red supergiant stars. The main reason is that an extra amount of energy is needed to expand a more compact structure.

Panel (b) in Fig. \ref{fig:general 7} illustrates the HR diagram of three models with different convective overshooting parameters in the binary system.
The convective overshoot is important to increase the luminosity on or after the main sequence and thereby stellar winds can also be increased. When the hydrogen starts
to become exhausted in the core, a thin hydrogen burning shell begins to take shape and moves outward. This leads
to a subsequent expansion and cooling of
the stellar envelope. The star B3 with a larger overshooting rapidly moves
to the redward in the HR diagram and becomes a red
supergiant in comparison with the model B1 with a smaller overshooting. This is the so called mirror effect.
The low temperature from the expansion significantly
increases the envelope opacity at the RSG phase, which can make it easier for photons to be caught by the
outer envelope, leading to increased radiation pressure and stronger line-driven winds.
Moreover, the outer envelopes are also less tightly bound, since the radius is
now several hundred solar radii or larger and the gravitational potential becomes shallow.
Both of these effects enormously reinforce
the mass loss and will result in a loss of a significant part of the outer
envelope.



During mass transfer phase via the RLOF, the stars lose most of their hydrogen-rich envelopes on a thermal time-scale. The mass-transfer phase proceeds at similar effective temperatures in three models. The main reason is that the size of the Roche lobe is approximately the same in three binaries. We have noted that the blue loop behavior is almost absent in all single stars in comparison with the binary system (cf., Panel a in Fig. \ref{fig:general 7}). We attribute to the presence of a thick hydrogen envelope in the single stars. This implies that there exists a threshold for the hydrogen envelope mass which can take the shape of the blue loop.

The inner layers with higher temperature can be exposed by RLOF. Actually, an evolution to the blue after the red supergiant appears only if the ratio of the core mass to the total mass can exceed a given limit of around $70\%$. This implies that the yellow or blue supergiant will
still have a hydrogen envelope that is comprised of a maximum of about $30\%$ of the total mass. The hottest point of the blue loop in model B1 corresponds to the minimum in the stellar radius when the core helium drops at $\rm Y_{c, He} \sim 0.25$, after which the envelope starts to expand and the star approaches the red giant branch again when the core helium $\rm Y_{c, He} \sim 0.02$.

We also see that the blue excursion can be triggered earlier in B3 with a larger overshooting than the B1 model with a smaller overshooting.
The reason is that the higher luminosity induced by larger convective overshooting can trigger stronger stellar winds after the RLOF. The donor star
loses a large amount of hydrogen during the blue excursion phase when the WR wind of \cite{Nugis2000} is adopted. The remaining hydrogen envelope can be completely eliminated by strong winds.
The binary model B3 will explode as a type $\rm \uppercase\expandafter{\romannumeral1}$b supernova because the hydrogen envelope is less than  $1 \%$ of the total mass.


\subsubsection{Metallicity effect}
Panel (c) in Fig. \ref{fig:general 6} shows the radius as a function of the evolutionary age at various metallicities. The radius is smaller at lower metallicity during the main sequence. There are two main reasons for this fact. Firstly, because an extremely low metallicity implies a weaker CNO-cycle, the nuclear reaction strongly depends on pp-chains to produce its nuclear energy at the beginning of its evolution. Since pp-chains are much less sensitive to temperature than CNO-cycle, the star has to contract more to obtain a higher central temperature. Secondly, the opacity is lower and thus the envelope is more
transparent at low metallicity. The radiative gradient $\rm \nabla_{rad}\propto\frac{\kappa P}{T^{4}}$ is lower and the star remains more compact.

Before RLOF, we can see that the model with lower metallicity can display a larger expansion compared to the one with higher metallicity. The reason is that
the residual hydrogen mass in the envelope is closely related to the stellar radius. The stellar radius generally decreases with the decreasing of the leftover hydrogen mass in the envelope. The progenitor radius of the type $\rm \uppercase\expandafter{\romannumeral2}$b SNe ranges form $\sim50  R_{\odot}$ of the BSG progenitor to $\sim 600  R_{\odot}$ of the RSG progenitor. One of the main parameters that can determine the final radius and
surface temperature of an SN IIb progenitor is the hydrogen
envelope mass at the pre-SN stage. We note that the mass range of the hydrogen envelope for type RSG SN
$\rm \uppercase\expandafter{\romannumeral2}$b is $\rm 0.35 M_{\odot} < M_{H} <0.5 M_{\odot}$, $\rm 0.15 M_{\odot} < M_{H} <0.35 M_{\odot}$ for type YSG SN $\rm \uppercase\expandafter{\romannumeral2}$b, and  $\rm \sim0.033 M_{\odot} < M_{H} <0.15 M_{\odot}$ for the type BSG SN $\rm \uppercase\expandafter{\romannumeral2}$b (cf., Table \ref{table1}). \cite{Gilkis2022} proposed that the type $\rm \uppercase\expandafter{\romannumeral1}$b and $\rm \uppercase\expandafter{\romannumeral2}$b hydrogen mass threshold is 0.033 $M_{\odot}$. This can be approximately consistent with
the previous estimates for the SNe $\rm \uppercase\expandafter{\romannumeral2}$b:
$\rm M_{H} = 0.2-0.4 M_{\odot}$ for SN 1993J and SN 2013df (RSG progenitor), $\rm M_{H} \simeq 0.1 M_{\odot}$ for SN 2011dh (YSG progenitor),
and $\rm M_{H} \simeq 0.06 M_{\odot}$ for SN 2008ax (BSG progenitor).

Stars with equal masses on the zero age main sequence but with different
initial metallicities have different pre-supernova structures for
a variety of reasons. Most importantly, if the amount of mass loss are very low in the single stars,
the pre-supernova star, including its helium core is larger. Finally, it has a larger compactness \cite{Sukhbold2014}.
We can see that the primary star in the binary system can deviate from this trend because the helium core can be reduced by RLOF.
Moreover, low metallicity implies a smaller initial helium mass fraction and
more hydrogen. The final helium core mass is sensitive to this
and is reduced accordingly (cf. Panel b in Fig. \ref{fig:general 3}).

Panel (c) in Fig. \ref{fig:general 7} illustrates the HR diagram of three models at various metallicities.
The stars with the low metallicity become bluer compared with the counterpart with high metallicity.
The central temperature is slightly higher at low metallicity because of the lower abundance of CNO which can catalyze the nuclear reaction.
When core hydrogen is exhausted, the star with low metallicity starts the central helium burning at the blue side of the HR diagram compared with the one with high metallicity.

Note that the primary star in three binaries can
experience a second RSG phase
when the helium becomes depleted in the core. Low metallicity reduces the energy
generation of the CNO cycle. The more active H-burning shell in the high
metallicity case reduces the pressure on the helium core. The metal-rich model is hotter and more compact than the metal-poor model at the core carbon exhaustion because the metal-poor model can retain more hydrogen envelope.
Therefore, low metallicity stars give rise to the slightly stripped Type $\rm \uppercase\expandafter{\romannumeral2}$b SNe, while
higher metallicity stars are prone to producing Type $\rm \uppercase\expandafter{\romannumeral1}$b or $\rm \uppercase\expandafter{\romannumeral1}$c SNe.

The hydrogen envelope in SN $\rm \uppercase\expandafter{\romannumeral2}$b progenitors becomes thicker for lower metallicity star, while the corresponding opacity becomes smaller. As these two effects compensate
each other, the radii of the $\rm \uppercase\expandafter{\romannumeral2}$b progenitor for both
low or high metallicity models B4 and B5 are found to be very similar (cf., Table 1).
The more compact supergiant envelope in model B5 places greater pressure on the helium core therein and can affect its subsequent evolution. It can determine if the progenitor is a red supergiant
or a yellow one. For example, the model B4 explodes as a red supergiants SN $\rm \uppercase\expandafter{\romannumeral2}$b progenitors
whereas model B5 ends as a YSG SN $\rm \uppercase\expandafter{\romannumeral2}$b progenitors.

\subsubsection{The orbital period effect}
Panel (d) in Fig. \ref{fig:general 6} shows radius as a function of the evolutionary age at various orbital periods.
The radius evolution becomes very distinct, with a monotonic increase of the radius for single stars until a maximum of $\rm \sim 10^{3} R_{\odot}$ is reached.
A non-monotonic increase of the radius occurs in the binary system because the thermal equilibrium is broken by the presence of RLOF. The more hydrogen stripped from the envelope, the larger the radius shrinks during the RLOF. When the mass transfer via RLOF goes down, the star gradually restores thermal equilibrium and its radius gradually increases. Actually, the radius can be greatly changed by the complication
of binary interaction: the presence of a thin hydrogen
envelope in some of the binary models leads to a more extended
envelope than in the corresponding single helium
star models in the absence of hydrogen envelope \citep{Yoon2017}.
The composition of the envelope also has an important impact on the evolution of the effective temperature.
The presence of hydrogen in the envelope of the extended progenitor can greatly reduce the effective temperature of the progenitor.
One can notice that single star progenitors tend to have larger radii than binary-star progenitors. For example, the radius of binary-star SN progenitor model B1 is $\rm 851 R_{\odot}$ while the radius of its single star counterpart S4 is $\rm 933 R_{\odot}$. There are two main reasons. Firstly, the stellar radius in binary systems is limited by the Roche lobe. Secondly, a smaller residual hydrogen envelope can be remained in the binary system due to mass transfer via the RLOF.

SNe $\rm \uppercase\expandafter{\romannumeral2}$b progenitors with smaller (larger)
envelope masses are more compact (extended) and hot
(cool). The ultra-stripped primary are believed to be very hot objects, emitting
the majority of their photons in the extreme ultraviolet and they can remain hidden by their companions.
The slightly stripped donors would be significantly cooler and more visible in their evolution \citep{Gotberg2017}.
A more progressive stripping process in an initial tighter system can cause the star to evolve toward the hydrogen-deficient WR star.
The mass-loss rate just before the SN explosion contains important information about their evolutionary paths.
The mass-loss property is reflected in the density of circum-stellar matter.
The radius of supernovae $\rm \uppercase\expandafter{\romannumeral2}$b covers the region form $\sim50  R_{\odot}$ of the BSG progenitor to $\sim 600  R_{\odot}$ of the RSG progenitor.
\cite{Maeda2015} have displayed that there is a close relationship
between the mass-loss rate of stellar winds and the progenitor radius
prior to the explosion. They presented that more
extended progenitors ($\sim 600 R_{\odot}$; e.g., 1993J, 2013df) have a
higher mass-loss rate that up to $\rm \sim 10^{-5} M_{\odot}/yr$, while less
extended progenitors ($\sim 200 R_{\odot}$; e.g., SN 2011dh) have a
moderate mass-loss rate ($\rm \sim3 \times 10^{-6} M_{\odot}/yr$).
\cite{Ouchi2017} interpreted that the less extended type $\rm \uppercase\expandafter{\romannumeral2}$b SN  progenitor has not only a smaller remanent envelope mass to transfer but also a larger value of equilibrium index $\xi_{eq}=(\frac{\partial \log R}{\partial \log M})_{eq}$ which denotes the variation in primary radius in response to the reduction of the envelope mass, assuming that this mass loss proceeds slowly enough to maintain the primary in thermal equilibrium. The larger value of $\rm \xi_{eq}$ indicates that the progenitor shrinks faster in response to the mass loss. However, \cite{van2005} reported that the mass loss rate via stellar winds can reach as high as $\sim 10^{-4}M_{\odot}/yr$ from some red supergiant stars. It may also be an alternative interpretation that the strong mass loss for the more extended progenitor can be understood by with a post-RLOF stellar wind of red supergiants in the model B1. The radius of the primary in model B1 can attain a value of $478 R_{\odot}$ and its stellar wind has a value of $\rm 2.58\times 10^{-6} M_{\odot}/yr$ (RSG progenitor of Supernovae $\rm \uppercase\expandafter{\romannumeral2}$b). These theoretical results are approximatively consistent with the observations of the type $\rm \uppercase\expandafter{\romannumeral2}$b SNe 1993J (cf. Table \ref{table2})\citep{Sravan2020}.

The thin envelope of SN $\rm \uppercase\expandafter{\romannumeral2}$b has a very
low mass compared to the envelope of an SN $\rm \uppercase\expandafter{\romannumeral2}$P, still soak up
a lot of energy from the SN shock, leading to the subsequent strong cooling
emission.
The large radius ($\rm \sim 600 R_{\odot}$) of SN 1993J is due to a thin extended
hydrogen envelope around the progenitor and should have important impact on
shock break-outs and bolometric light curves.
The extended model gives rise to a apparent peak at the early stage (at the time of 5 days after the SN explosion) of the observed bolometric light curves of the SN $\rm \uppercase\expandafter{\romannumeral2}$b while the compact progenitor shows a much smaller bulge. The main difference is due to the extra amount of energy required to expand a more compact structure \citep{Bersten2012}.

Panel (d) in Fig. \ref{fig:general 7} illustrates the evolution of several models with various initial orbital periods in the HR diagram.
Models with different initial orbital periods can give rise to the blue, yellow, and
red supergiants SN $\rm \uppercase\expandafter{\romannumeral2}$b progenitors. Whether a star explodes as red or blue supergiant, and how much hydrogen envelope remains depending on the mass loss via stellar winds or the RLOF in the binary system.
The extended blue loop can be greatly influenced by the enhanced mass-loss via RLOF. The effective temperature gradually reduces with the increase of the size of the radius and the initial orbital period.
The primary in an initial tighter system is hotter and more compact than the one in a wider system
because an initial wider system can retain more hydrogen in its envelope, which allows it to expand greatly. The position in the HRD of the endpoint of the evolution heavily depends on the leftover mass of the hydrogen envelope. If the hydrogen envelope contains more than about $5\%$ of the initial mass, the star evolves into a red supergiant star with a lower effective temperature at the pre-supernova stage \citep{Meynet2015}. If the hydrogen envelope keeps less than  $1 \%$ of the total mass, the star becomes a blue supergiant or a Wolf-Reyet star. The primary star in the wide system remains a significant hydrogen layer after RLOF which can sustain a hydrogen burning shell while the one in the tight system loses its hydrogen-rich envelope earlier due to an ultra-stripped RLOF. The hydrogen shell burning governs the nuclear luminosity during the time of helium exhaustion and this results in a local maximum in the radius. The inflated
hydrogen envelope of supergiant progenitor has important impact on the early-time light curves of SNe $\rm \uppercase\expandafter{\romannumeral2}$b during
the shock-cooling phase. An inflated hydrogen envelope can contribute to sustained high luminosity.

The wind mass loss of a single star with mass $\rm< 20 M_{\odot}$ is not strong enough to completely
expel the envelope (cf., panel a in Fig. \ref{fig:general 5}). Instead, for a less massive star ($\rm < 20 M_{\odot}$) to become a stripped envelope supernova, a significant mass transfer in a close binary system is needed. In this
case, the stripped envelope supernova has lower mass. RLOF can avoid the redward evolution after the main sequence in the initial tighter systems B6 and B7.

The pre-explosion photometry of the
progenitor is critical for the evolution and final appearance of an SN $\rm \uppercase\expandafter{\romannumeral2}$b
and its companion star. To date, a total of five type $\rm \uppercase\expandafter{\romannumeral2}$b SN progenitors have been identified in pre-explosion imaging, from SN 1993J to the more recent SN 2016gkg. Some observational evidence suggests the presence of its companion near the progenitor stars of SN1993J, SN2001ig, and SN2011dh. The observational position of two component stars in the HR diagram can also provide us with a good target to construct the theoretical model and the evolutionary properties. At the time of SN explosion, the primary star in model B1 fits well with the the pre-explosion observations. However, the location of the companion star of SN 1993J in the HR diagram is in approximate agreement with observations. The luminosity of the secondary practically match the observations, but the star is too blue (cf., panel f in Fig. \ref{fig:general 7}). The secondary in our model is shifted to a higher effective temperature by just $\rm \Delta \log T_{eff}\simeq 0.14$. \cite{Stancliffe2009} have claimed that it is extremely difficult for the secondary star to attain the observational position in the HR diagram. The position indicates that the secondary star appears an overluminous B supergiant and is
extremely close to (or just beyond) the right side of its main sequence band. Its position predicted by the theoretical model is extremely sensitive to the value of the accretion efficiency $\rm \beta$ \citep{Benvenuto2013}. Both the effective temperature and luminosity of the companion star of SN 1993J can be reduced by a smaller accretion efficiency. They proved that this can only be done in a very narrow range in initial masses and periods. \cite{Stancliffe2009} have presented that if mass transfer is conservative, the right location of the secondary in the HR diagram can be reproduced by a system consisting of a 15 $\rm M_{\odot}$ primary and a 14 $\rm M_{\odot}$ secondary in an orbit with an initial period of 2100 days.  However, they made using of a high metallicity model with $\rm Z=0.04$. We simulate the system of SN 1993J is comprised of a 16 $\rm M_{\odot}$ primary and a 15 $\rm M_{\odot}$ secondary with an initial orbital period of 1100 days. The metallicity is $\rm Z=0.04$ and the accretion efficiency is adjusted to a value of $\rm \beta=0.15$. The final evolutionary positions of two components in the HR diagram are in good agreement with the observations (cf., Table \ref{table2} and panel f in Fig. \ref{fig:general 7}). In the model B12, the luminosity and the effective temperature of the primary have the values of $\rm \log L/L_{\odot}=4.91$ and $\rm \log T_{eff}=3.88$, respectively.
The observations of the SN 2008ax progenitor implies that the luminosity and the effective temperature of the primary are $\rm \log L/L_{\odot}=4.42-5.3$ and $\rm \log T_{eff}=3.88-4.3$, respectively \citep{Sravan2020}.
These theoretical results can be also consistent with the observations of SN 2008ax (BSG progenitor of Supernovae $\rm \uppercase\expandafter{\romannumeral2}$b).


\subsection{Carbon profile at the core carbon exhaustion}
\subsubsection{The overshooting effect}

Panel (a) in Fig. \ref{fig:general 8} shows the mass fraction of carbon as a function of the lagrangian mass for the non-rotating and rotating single stars at the core carbon exhaustion.
The carbon-oxygen core mass which scales with the mass of helium core increases with the initial mass of the
star at all metallicities.  For example, the CO core mass is 2.853 $M_{\odot}$ for model S4 while it is  3.418 $M_{\odot}$ for model S7. One can see that the  overshooting governs the final mass of the carbon-oxygen core. A higher overshooting parameter can result in a
larger final carbon-oxygen core masse. The growth of the helium-burning core overshooting can bring about an additional supply of helium above the core that favors the formation of a larger CO core at the core helium exhaustion. Actually, all single stars and the peeled primary stars in our calculations display similar density profiles in the iron core (i.e., $\rm M < \sim 1.6 M_{\odot}$). However, the star shows a shallower drop of the density profile above the iron core for the more massive helium core. The local density tends to be higher while the central core has a higher temperature.

An important effect is that the increase of overshooting leads to a lower ratio of carbon to oxygen at the end of helium
core burning, which can have strong impact on the strength of subsequent carbon
burning and the final size of the iron core. For a high carbon abundance in the model with smaller overshooting parameter, the phase of convective
carbon shell burning may proceed longer, typically leading to smaller
carbon-exhausted cores. This outcome in turn produces smaller iron
cores with steeper density gradient outside the iron core and results in a pre-SN structure that more easily produces a successful supernova \citep{Limongi2018}. Furthermore, the formation of a larger convective core induced by the overshooting slows down the
contraction of the core as well as its heating.

It can be noticed that at the mass coordinate of $1.0 M_{\odot}$ the central carbon mass fraction reduces with increasing of overshooting parameter.
Generally, the mass fraction $\rm ^{12}C$ left by the core helium
burning decreases with the CO core mass. This is because the star with a larger convective overshooting has a higher reference core mass.
This condition favors the rate of alpha captures onto carbon over the triple-alpha process. Therefore, carbon is destroyed more efficiently in the larger core induced by the convective overshooting and the mass fraction of oxygen
is higher in the enlarged core.
This contributes to a faster outward shift of the carbon burning shell and a more compact core of the star \cite{Chieffi2013}.

\subsubsection{The metallicity effect}
Panel (b) in Fig. \ref{fig:general 8} shows the mass fraction of carbon as a function of the lagrangian mass in nonrotating binary models with different metallicities at the end of the core carbon-burning phase.
Actually, mass loss in the more massive star is strong enough to reduce substantially
the total mass and therefore to reduce the hydrogen convective core during
the main sequence. As a consequence, the helium core mass is smaller than it would be in the case of weak mass loss.
Therefore, a strong mass loss in the single star with high metallicity will drive evolution
toward a smaller carbon-oxygen core in contrast to the model with low metallicity.

However, RLOF can make the low metallicity primary star B4 transfer more hydrogen envelopes to the companion.
Therefore, the carbon-oxygen core mass in the low metallicity star becomes smaller due to RLOF.
The star tends to behave as a lower mass star and evolves toward lower central temperature and higher central density. Such an occurrence has
five outcomes. Firstly, the helium convective core shrinks progressively in mass and
leaves a zone with variable chemical composition. Secondly, the lifetime of core helium burning
goes up accordingly. Thirdly, the total luminosity progressively falls down and then the star shifts downward in the HR diagram. Fourth, the local density above the iron core tends to be higher in an initial high metallicity circumstance while the central temperature becomes larger. The mass fraction of $\rm ^{12}C$ at the time of core helium exhaustion becomes
larger than it would be in high metallicity binary system. Finally, the CO core at core He exhaustion is
smaller than it would be in high metallicity binary system.
Generally speaking, stellar compact at the pre-supernova stage increases with the carbon-oxygen mass. Therefore, it is found that the primary star with low metallicity is easier to explode. As a consequence, we expect these models to produce smaller remnant
masses and eventually, to give rise to faint and failed supernovae.

\subsubsection{The orbital period effect}
Panel (c) in Fig. \ref{fig:general 8} shows the carbon profiles for the primary star with different orbital periods in the binary system at the time of core carbon exhaustion. For a given initial mass, the final CO core mass may origin from different initial orbital periods. As can be seen a smaller carbon-oxygen core can be induced by the RLOF in an initial tighter system B6. The local density above the iron core tends to be smaller in this system while the central temperature becomes lower.

We also find that a higher central carbon abundance can be reached in this system. A hydrogen-burning shell can be extinguished because of RLOF. This leads to a smaller helium core and a higher ratio of carbon to oxygen at the end of carbon core burning,
which have an important impact on the strength of subsequent carbon burning and the
final size of the iron core \citep{Brown2001}.
The bolometric luminosity of the SN progenitor can mainly be resolved by helium shell burning, which is in turn largely resolved by the mass of the CO core.
However, the strength of the secondary
carbon convective shell progressively weakens because the fraction of carbon left after the core helium exhaustion inversely scales with the CO core mass.
The final degree of core compactness
of a star can be increased because it heavily depends on the formation and development of the various carbon convective episodes \citep{Chieffi2000,Limongi2009}.


\subsubsection{The effect of rotation}
Panel (d) in Fig.            \ref{fig:general 8} illustrates the carbon profile of the single star and the primary star in the binary system for different rotational velocities at the time of core carbon exhaustion.
Rapid rotation can increase the mass of helium and carbon-oxygen cores in both single stars and binary systems. The greater the initial rotational speed, the greater the helium and carbon-oxygen core mass. The corresponding central density tends to be higher.
Therefore, rotational mixing can significantly reduce the mass fraction $\rm ^{12}C$ at the core helium
exhaustion. Rotation can also continuously mix the helium outside the helium burning region and give rise to more $\rm ^{12}C$.
At a given initial rotational velocity, the mass of carbon-oxygen core in single stars is larger than in a binary system.
For example, the carbon-oxygen core mass is 3.16 $M_{\odot}$ for model B11 while it is  2.62 $M_{\odot}$ for model S6.
The main reason is that RLOF can extinguish the hydrogen burning shell which may contribute an amount of helium to the helium core.
For a given initial mass, rotating models behave like more massive stars and therefore they end their life
with more compact structures.
\setlength{\tabcolsep}{0.5mm}{ 
\setlength{\LTcapwidth}{17.0in}   
\begin{center}
\scriptsize{
\begin{longtable}{lcccccccccccccccccccccc}
\caption{\label{table3}Major evolutionary parameters for nine models including single stars and the primary star in binaries.}\\
   \hline \hline
Sequence &Age & $M_1$& $\rm \log(\frac{R}{R_{\odot}})$ &$\log T_{\rm eff}$& $\log (\frac{L}{L_{\odot}})$&$\log T_{\rm c}$&$\log \rho_{\rm c}$& $\rm \frac{N}{N_{\rm ini}}$& $V_{\rm eq}$& $X_{\rm H}$& $Y_{\rm He}$ & $\log(^{12}C)$& $\log(^{14}N)$& $\log(^{16}O)$& $\rm \frac{^{14}N}{^{12}C}$\\\hline
  &Myr&$M_{\odot}$& & K&  &K&$\rm g/cm^{3}$&&km/s \\
    \hline
\endfirsthead

  \hline
Sequence,&Age & $M_1$ & $\rm \log(\frac{R}{R_{\odot}})$ &$\log T_{\rm eff}$& $\log (\frac{L}{L_{\odot}})$&$\log T_{\rm c}$& $\log \rho_{\rm c}$& $\rm \frac{N}{N_{\rm ini}}$ & $V_{\rm eq}$& $X_{\rm H}$& $Y_{\rm He}$ & $\log(^{12}C)$& $\log(^{14}N)$& $\log(^{16}O)$& $\rm \frac{^{14}N}{^{12}C}$\\
    \hline
    \endhead
    \hline
    \endfoot
ZAMS&&&&&&&&&\\

S1&0.103&15.000&0.694&4.487&4.290&7.534&0.787&0.995&0.000&0.700&0.280&-2.463&-2.996&-2.029&0.293\\
S4&0.144&17.000&0.726&4.510&4.447&7.543&0.734&0.993&0.000&0.700&0.280& -2.463&-2.996&-2.029&0.293\\
S6&0.169&17.000&0.768&4.483&4.422&7.541&0.741&1.000&400.00&0.700&0.280&-2.463&-2.996&-2.029&0.293\\
B1&0.104&16.999&0.725&4.511&4.446&7.543&0.737&1.000&0.000&0.700&0.280&-2.463&-2.996&-2.029&0.293\\
B3&0.174&16.998&0.726&4.510&4.448&7.543&0.734&1.000&0.000&0.700&0.280&-2.463&-2.996&-2.029&0.293\\
B4&0.043&17.000&0.698&4.512&4.397&7.560&0.811&1.000&0.000&0.736&0.256&-2.861&-3.394&-2.427&0.293\\
B6&0.107&16.999&0.725&4.511&4.446&7.543&0.737&1.000&0.000&0.700&0.280&-2.463&-2.996&-2.029&0.293\\
B9&0.104&16.999&0.725&4.511&4.446&7.543&0.737&1.000&0.000&0.700&0.280&-2.463&-2.996&-2.029&0.293\\
B11&0.199&16.997&0.769&4.483&4.423&7.540&0.740&1.000&400.00&0.700&0.280&-2.463&-2.996&-2.029&0.293\\

\hline
ECHB&&&&&&&&&\\

S1&11.642&14.619&1.075&4.391&4.670&7.732&1.301&1.000&0.000&0.700&0.280&-2.463&-2.996&-2.029&0.293\\
S4&9.900&16.506&1.124&4.405&4.820&7.743&1.257&1.000&0.000&0.700&0.280&-2.463&-2.996&-2.029&0.293\\
S6&12.841&15.993&1.105&4.445&4.943&7.751&1.224&8.488&131.725&0.596&0.384&-3.224&-2.068&-2.344&14.328\\
B1&9.923&16.455&1.119&4.406&4.816&7.744&1.265&1.000&0.000&0.700&0.280&-2.463&-2.996&-2.029&0.293\\
B3&11.136&16.069&1.315&4.332&4.913&7.759&1.271&1.000&0.000&0.700&0.280&-2.463&-2.996&-2.029&0.293\\
B4&11.133&16.797&1.075&4.424&4.800&7.864&1.776&1.000&0.000&0.736&0.256&-2.861&-3.394&-2.427&0.293\\
B6&10.629&7.479&1.148&4.307&4.478&7.746&1.457&5.712&0.000&0.691&0.288&-4.487&-2.240&-2.072&176.912\\
B9&9.923&16.455&1.119&4.406&4.816&7.744&1.265&1.000&0.000&0.700&0.280&-2.463&-2.996&-2.029&0.293\\
B11&11.808&15.774&1.158&4.402&4.877&7.746&1.238&6.529&72.147&0.654&0.326&-3.037&-2.182&-2.197&7.167\\

\hline
BCHEB&&&&&&&&&\\

S1&11.642&14.619&1.075&4.391&4.670&7.732&1.301&1.000&0.000&0.700&0.280&-2.463&-2.996&-2.029&0.293\\
S4&9.905&16.504&1.102&4.422&4.845&7.816&1.525&1.000&0.000&0.700&0.280&-2.463&-2.996&-2.029&0.293\\
S6&12.846&15.993&1.075&4.467&4.971&7.839&1.534&8.488&144.644&0.596&0.384&-3.224&-2.068&-2.344&14.331\\
B1&9.941&16.449&1.385&4.286&4.866&8.038&2.548&1.000&0.000&0.700&0.280&-2.463&-2.996&-2.029&0.293\\
B3&11.148&16.061&1.586&4.202&4.934&8.054&2.384&1.000&0.000&0.700&0.280&-2.463&-2.996&-2.029&0.293\\
B4&11.145&16.795&1.329&4.302&4.820&8.053&2.633&1.000&0.000&0.736&0.256&-2.861&-3.394&-2.427&0.293\\
B6&10.659&5.554&1.392&4.240&4.696&8.042&2.902&12.422&0.000&0.300&0.681&-4.037&-1.902&-3.176&136.285\\
B9&9.941&16.449&1.385&4.285&4.866&8.038&2.549&1.000&0.000&0.700&0.280&-2.463&-2.996&-2.029&0.293\\
B11&11.827&15.763&1.663&4.159&4.917&8.063&2.536&6.530&31.258&0.654&0.326&-3.037&-2.182&-2.197&7.171\\

\hline
ECHEB&&&&&&&&&\\

S1&12.788&12.926&2.881&3.541&4.878&8.905&5.430&3.943&0.000&0.645&0.334&-2.699&-2.401&-2.103&1.987\\
S4&10.988&13.447&2.880&3.548&4.907&8.478&3.587&4.334&0.000&0.632&0.348&-2.722&-2.360&-2.120&2.305\\
S6&13.761&9.950&2.956&3.582&5.192&8.911&5.297&10.065&0.077&0.467&0.513&-3.544&-1.994&-2.505&35.495\\
B1&10.924&5.177&2.687&3.665&4.987&8.906&5.307&9.619&0.000&0.478&0.502&-3.984&-2.013&-2.411&93.370\\
B3&11.918&5.442&0.194&4.943&5.113&8.922&5.232&12.493&0.000&0.000&0.981&-3.567&-1.900&-3.461&46.424\\
B4&12.153&5.011&2.712&3.639&4.933&8.902&5.343&9.985&0.000&0.526&0.466&-4.351&-2.395&-2.860&90.439\\
B6&12.000&3.347&0.626&4.627&4.715&8.885&5.486&12.546&0.000&0.000&0.981&-3.626&-1.898&-3.480&53.502\\
B9&10.909&6.848&2.955&3.533&4.993&8.905&5.318&5.874&0.000&0.584&0.396&-2.851&-2.228&-2.190&4.207\\
B11&12.668&5.202&0.234&4.913&5.074&8.918&5.257&12.353&95.589&0.000&0.975&-2.289&-1.905&-3.257&2.424\\

\hline
BCCB&&&&&&&&&\\

S1&12.788&12.926&2.881&3.541&4.878&8.905&5.430&3.943&0.000&0.645&0.334&-2.699&-2.401&-2.103&1.987\\
S4&11.013&13.327&2.982&3.532&5.046&8.807&5.148&4.978&0.000&0.615&0.365&-2.780&-2.299&-2.146&3.023\\
S6&13.761&9.950&2.956&3.582&5.192&8.911&5.297&10.065&0.077&0.467&0.513&-3.544&-1.994&-2.505&35.495\\
B1&10.924&5.177&2.687&3.665&4.987&8.906&5.307&9.619&0.000&0.478&0.502&-3.984&-2.013&-2.411&93.370\\
B3&11.918&5.442&0.194&4.943&5.113&8.922&5.232&12.493&0.000&0.000&0.981&-3.567&-1.900&-3.461&46.424\\
B4&12.153&5.011&2.712&3.639&4.933&8.902&5.343&9.985&0.000&0.526&0.466&-4.351&-2.395&-2.860&90.439\\
B6&12.000&3.347&0.626&4.627&4.715&8.885&5.486&12.546&0.000&0.000&0.981&-3.626&-1.898&-3.480&53.502\\
B9&10.909&6.848&2.955&3.533&4.993&8.905&5.318&5.874&0.000&0.584&0.396&-2.851&-2.228&-2.190&4.207\\
B11&12.668&5.202&0.234&4.913&5.074&8.918&5.257&12.353&95.589&0.000&0.975&-2.289&-1.905&-3.257&2.424\\

\hline
ECCB&&&&&&&&&\\

S1&12.789&12.922&2.885&3.540&4.884&8.933&5.941&3.977&0.000&0.644&0.336&-2.701&-2.397&-2.104&2.015\\
S4&11.015&13.316&2.976&3.533&5.038&8.944&5.974&4.991&0.000&0.614&0.366&-2.781&-2.298&-2.146&3.039\\
S6&13.761&9.947&2.953&3.581&5.182&8.949&5.851&10.083&0.082&0.465&0.516&-3.546&-1.993&-2.508&35.715\\
B1&10.925&5.169&2.688&3.667&4.997&8.938&5.850&9.630&0.000&0.477&0.503&-3.983&-2.013&-2.413&93.333\\
B3&11.919&5.441&0.224&4.930&5.119&8.917&5.749&12.493&0.000&0.000&0.981&-3.567&-1.900&-3.461&46.424\\
B4&12.154&5.002&2.713&3.641&4.944&8.938&5.866&10.013&0.000&0.525&0.467&-4.350&-2.394&-2.864&90.444\\
B6&12.003&3.341&0.720&4.581&4.717&8.928&6.012&12.546&0.000&0.000&0.981&-3.625&-1.898&-3.481&53.334\\
B9&10.910&6.833&2.956&3.535&5.006&8.937&5.846&5.899&0.000&0.583&0.397&-2.854&-2.226&-2.191&4.250\\
B11&12.669&5.200&0.268&4.898&5.081&8.923&5.760&12.348&83.027&0.000&0.975&-2.287&-1.905&-3.256&2.408\\

\hline
BCNEB&&&&&&&&&\\

S1&12.790&12.919&2.875&3.542&4.870&9.178&7.165&3.984&0.000&0.644&0.336&-2.701&-2.396&-2.105&2.020\\
S4&11.015&13.316&2.977&3.533&5.038&9.181&6.849&4.991&0.000&0.614&0.366&-2.781&-2.298&-2.146&3.039\\
S6&13.761&9.946&2.951&3.583&5.187&9.178&6.852&10.093&0.083&0.463&0.517&-3.547&-1.992&-2.509&35.827\\
B1&10.925&5.165&2.682&3.664&4.974&9.179&7.071&9.636&0.000&0.477&0.503&-3.982&-2.013&-2.413&93.301\\
B3&11.919&5.440&0.245&4.917&5.113&9.180&6.679&12.493&0.000&0.000&0.981&-3.567&-1.900&-3.461&46.424\\
B4&12.155&4.998&2.711&3.639&4.931&9.182&7.128&10.032&0.000&0.525&0.467&-4.349&-2.393&-2.867&90.437\\
B6&..&..&..&..&..&..&..&..&..&..&..&..&..&..&..\\
B9&10.910&6.825&2.954&3.534&4.996&9.180&7.048&5.915&0.000&0.583&0.397&-2.856&-2.225&-2.192&4.277\\
B11&12.669&5.199&0.303&4.878&5.071&9.125&6.804&12.346&71.602&0.000&0.975&-2.285&-1.905&-3.256&2.401\\

\hline
ECNEB&&&&&&&&&\\

S1&12.790&12.919&2.876&3.541&4.870&9.284&6.968&3.984&0.000&0.644&0.336&-2.701&-2.396&-2.105&2.020\\
S4&11.015&13.316&2.978&3.532&5.038&9.302&6.772&4.991&0.000&0.614&0.366&-2.781&-2.298&-2.146&3.039\\
S6&13.761&9.946&2.953&3.582&5.189&9.309&6.682&10.093&0.082&0.463&0.517&-3.547&-1.992&-2.509&35.827\\
B1&10.925&5.165&2.684&3.664&4.979&9.275&7.048&9.636&0.000&0.477&0.503&-3.982&-2.013&-2.413&93.301\\
B3&11.919&5.440&0.282&4.895&5.098&9.265&6.968&12.493&0.000&0.000&0.981&-3.567&-1.900&-3.461&46.424\\
B4&12.155&4.998&2.713&3.639&4.936&9.262&6.924&10.032&0.000&0.525&0.467&-4.349&-2.393&-2.867&90.437\\
B6&..&..&..&..&..&..&..&..&..&..&..&..&..&..&..\\
B9&10.910&6.825&2.954&3.534&4.996&9.180&7.048&5.915&0.000&0.583&0.397&-2.856&-2.225&-2.192&4.277\\
B11&12.669&5.199&0.326&4.866&5.070&9.310&7.705&12.346&71.202&0.000&0.975&-2.285&-1.905&-3.256&2.401\\

\hline
BROLF1&&&&&&&&&\\

B1&9.955&16.442&2.368&3.789&4.843&8.195&2.928&1.000&0.000&0.700&0.280&-2.463&-2.996&-2.029&0.293\\
B3&11.152&16.058&2.371&3.794&4.869&8.168&2.761&1.000&0.000&0.700&0.280&-2.463&-2.996&-2.029&0.293\\
B4&11.164&16.789&2.363&3.775&4.778&8.218&3.024&1.000&0.000&0.736&0.256&-2.861&-3.394&-2.427&0.293\\
B6&8.534&16.835&1.037&4.424&4.724&7.585&0.768&1.000&0.000&0.700&0.280&-2.463&-2.996&-2.029&0.293\\
B9&9.966&16.423&2.840&3.531&4.757&8.211&2.961&2.176&0.000&0.685&0.294&-2.563&-2.659&-2.048&0.801\\
B11&11.832&15.760&2.378&3.783&4.843&8.164&2.857&6.530&5.519&0.654&0.326&-3.037&-2.182&-2.197&7.174\\

\hline
EROLF1&&&&&&&&&\\

B1&9.964&5.594&2.671&3.618&4.768&8.206&2.945&9.575&0.000&0.479&0.501&-3.915&-2.015&-2.408&79.428\\
B3&11.159&6.820&2.574&3.751&5.104&8.204&2.807&9.441&0.000&0.514&0.466&-3.766&-2.021&-2.398&55.484\\
B4&11.174&5.372&2.688&3.598&4.720&8.224&3.025&9.866&0.000&0.528&0.464&-4.294&-2.400&-2.846&78.234\\
B6&10.493&7.500&1.215&4.260&4.425&7.636&1.118&5.712&0.000&0.691&0.288&-4.487&-2.240&-2.072&176.911\\
B9&9.997&10.171&2.887&3.510&4.766&8.230&2.999&4.372&0.000&0.626&0.354&-2.714&-2.356&-2.126&2.280\\
B11&11.846&6.866&2.614&3.718&5.054&8.208&2.893&11.019&0.385&0.375&0.605&-3.609&-1.954&-2.683&45.107\\

\hline
BROLF2&&&&&&&&&\\

B1&10.916&5.241&2.665&3.664&4.940&8.706&4.682&9.589&0.000&0.479&0.502&-3.960&-2.015&-2.408&88.108\\
B3&..&..&..&..&..&..&..&..&..&..&..&..&..&..&..\\
B4&12.134&5.194&2.684&3.625&4.819&8.621&4.397&9.868&0.000&0.528&0.464&-4.295&-2.400&-2.846&78.485\\
B6&10.635&7.478&1.215&4.283&4.515&7.811&1.791&5.712&0.000&0.691&0.288&-4.487&-2.240&-2.072&176.912\\
B9&10.519&8.940&2.888&3.509&4.765&8.286&3.077&4.372&0.000&0.626&0.354&-2.714&-2.356&-2.126&2.280\\
B11&..&..&..&..&..&..&..&..&..&..&..&..&..&..&..\\

\hline
EROLF2&&&&&&&&&\\

B1&..&..&..&..&..&..&..&..&..&..&..&..&..&..&..\\
B3&..&..&..&..&..&..&..&..&..&..&..&..&..&..&..\\
B4&..&..&..&..&..&..&..&..&..&..&..&..&..&..&..\\
B6&10.680&3.976&1.553&4.167&4.727&8.179&3.126&12.543&0.000&0.169&0.811&-3.927&-1.898&-3.309&106.831\\
B9&..&..&..&..&..&..&..&..&..&..&..&..&..&..&..\\
B11&..&..&..&..&..&..&..&..&..&..&..&..&..&..&..\\
\hline
EOC&&&&&&&&&\\

S1&12.790&12.919&2.876&3.541&4.869&9.301&7.569&3.984&0.000&0.644&0.336&-2.701&-2.396&-2.105&2.020\\
S4&11.015&13.316&2.977&3.532&5.039&9.301&7.473&4.991&0.000&0.614&0.366&-2.781&-2.298&-2.146&3.039\\
S6&13.761&9.946&2.956&3.580&5.186&9.735&8.990&10.093&0.082&0.463&0.517&-3.547&-1.992&-2.509&35.827\\

B1&10.925&5.165&2.685&3.661&4.968&9.933&9.652&9.636&0.000&0.477&0.503&-3.982&-2.013&-2.413&93.301\\
B3&11.919&5.440&0.320&4.869&5.070&9.956&9.908&12.493&0.000&0.000&0.981&-3.567&-1.900&-3.461&46.424\\
B4&12.155&4.998&2.714&3.638&4.935&9.953&9.765&10.032&0.000&0.525&0.467&-4.349&-2.393&-2.867&90.437\\
B6&12.005&3.339&0.818&4.519&4.665&8.979&7.211&12.545&0.000&0.000&0.981&-3.624&-1.898&-3.481&53.243\\
B9&10.910&6.825&2.955&3.532&4.992&9.940&9.626&5.915&0.000&0.583&0.397&-2.856&-2.225&-2.192&4.277\\
B11&12.669&5.199&0.420&4.803&5.005&9.496&8.485&12.346&57.367&0.000&0.975&-2.285&-1.905&-3.256&2.401\\

  & & & & & &  & & & &\\
\hline\hline
\end{longtable}
\begin{tablenotes}
\footnotesize
\item[1)] Abbreviations: ZAMS-zero age main sequence; ECHB-the terminal of core hydrogen burning; ECHEB-the end of core helium burning; ECCB-the end of core carbon burning; BCNEB-the beginning of core neon burning;
ECNEB-the end of core neon burning; BROLF1-the beginning of the first episode of RLOF; EROLF1-the end of the first episode of RLOF; BROLF2-the beginning of the second episode of RLOF; EROLF2-the end of the second episode of RLOF; BROLF3-the beginning of the third episode of RLOF; EROLF3-the end of the third episode of RLOF; EOC-the end of the calculation.
\end{tablenotes}
}
\end{center}
\section{Discussion}
 {There is an unsolved, long-standing problem that the observed rates of type $\rm\uppercase\expandafter{\romannumeral2}$b SNe seem to be much higher than the one which is predicted by binary evolution, in particular type $\rm\uppercase\expandafter{\romannumeral2}$b SNe with red/yellow supergiant progenitors. \cite{Claeys2011} have presented that binary evolution predicts roughly $0.6 \%$ of all core collapse SNe to be type $\rm \uppercase\expandafter{\romannumeral2}$b SNe, about a factor 5 lower than the observed rate. Actually, they restrained the parameter spaces at solar metallicity to the initial primary mass of 15 $\rm M_{\odot}$,  initial secondary masses 10-15 $\rm M_{\odot}$, initial orbital periods 800-2100 days.  Because of their limited parameter space coverage, they were not able to obtan robust relative rates. Another limitation is that they restrict their analysis to progenitors that explode with 0.1-0.5 $\rm M_{\odot}$ of residual hydrogen envelope. The low mass limit of 0.1 $\rm M_{\odot}$ excludes the group of more compact SN $\rm \uppercase\expandafter{\romannumeral2}$b progenitors obtained from theoretical models and it can be significantly reduced by the recent observations.}

 {Actually, this fraction increases if the companion star can accrete only a small fraction of the transferred mass via RLOF, and if the mass outflow carries relatively low angular momentum. If more material can escape from the binary system and take along its orbital momentum, this will cause the orbit to shrink faster, promoting  parameter space for the evolution of contact or unstable mass transfer. This evolutionary pathway might increase the probability of the CEE channel to produce type $\rm \uppercase\expandafter{\romannumeral2}$b SNe and has not included in the calculation of \cite{Claeys2011}. A smaller accretion efficiency can broaden the orbit and therefore the Roche lobe. This results in smaller mass-transfer rates via the RLOF and therefore larger hydrogen envelope masses of the donor at the time of explosion. A smaller accretion efficiency tends to generate the stable mass-transfer and it leads to more type $\rm \uppercase\expandafter{\romannumeral2}$b SNe than type $\rm \uppercase\expandafter{\romannumeral1}$b SNe. Therefore, lower mass transfer efficiencies are also favorable for the production of SNe $\rm \uppercase\expandafter{\romannumeral2}$b with red/yellow supergiant progenitors.}

 {Moreover, if post-RLOF wind is lower than usually, it can greatly increase the rates that are predicted by binary evolution. \cite{Gilkis2019} commented that the wind mass-loss rate derived by \cite{Vink2017}, instead of the rate by \cite{Nugis2000}, greatly shifts binary progenitor models for core collapse SNe over a large initial parameter space from type $\rm \uppercase\expandafter{\romannumeral1}$b to type $\rm \uppercase\expandafter{\romannumeral2}$b. The mass-loss rate is expected to be smaller at lower metallicites and thus \cite{Yoon2017} presented that there would be even more type $\rm \uppercase\expandafter{\romannumeral2}$b SNe relative to type $\rm \uppercase\expandafter{\romannumeral1}$b SNe.}

 {In this paper, we intend to expand the parameter space for type $\rm \uppercase\expandafter{\romannumeral2}$b SN progenitors using detailed binary evolutionary calculations. With a smaller overshooting parameter, the stars have smaller core masses and hence are less luminous, with smaller radii. However, the final remnant masses tend to be similar to their counterpart with a larger overshooting parameter. As a result, these stars have much smaller final helium core masses and they lose less of their envelopes accordingly, retaining larger amounts of hydrogen at the point of explosion. This evolution may be in favor of the formation of SNe $\rm \uppercase\expandafter{\romannumeral2}$b with red/yellow.}

 {Rotation is thought to play a critical role in massive star evolution and rapidly rotating stars even are predicted to evolve chemically homogeneously due to the high efficiency of rotational mixing \citep{Maeder1987, de2009, Song2016}. The star evolves blueward without experiencing the RSG phase. This evolutionary pathway greatly reduces the mass of the hydrogen envelope and significantly increases the production rate of type $\rm \uppercase\expandafter{\romannumeral2}$b or $\rm \uppercase\expandafter{\romannumeral1}$b supernovae. Moreover,
rotationally enhanced mass-loss rates can also reduce the minimum mass required for a single star to remove its hydrogen envelope \citep{Meynet2003}. Mass loss by line-driven winds is closely related to the chemical abundances and the luminosity. Therefore, rotation mixing can enhance the surface helium fraction and the luminosity and this also increase the rate of SN $\rm \uppercase\expandafter{\romannumeral2}$b with red/yellow progenitors from single stars.}

 {\cite{Sravan2018} have found that it is very difficult to account for the rate of type $\rm \uppercase\expandafter{\romannumeral2}$b SNe at solar metallicity. They take a ratio of the type $\rm \uppercase\expandafter{\romannumeral2}$b SNe of about 10-12 $\%$ in high metallicity stellar populations and about 20 $\%$ in low metallicity populations. Therefore, the parameter space for binary SNe $\rm \uppercase\expandafter{\romannumeral2}$b rapidly increases  with the decreasing of metallicity. This is because of evolutionary channels to SNe $\rm \uppercase\expandafter{\romannumeral2}$b via Case early-B mass transfer that is only viable at low metallicity. In brief, a new statistical investigation is needed to compare binary models with the overall rates of different types of core collapse SNe in the future work.}

\section{Conclusion and summary}
        In this paper, we investigate the evolution of stars that lost most of their hydrogen-rich envelope because of the interaction with their companions. We consider the binary systems with various overshooting, metallicities, initial orbital periods, and initial rotational velocities and these physical factors are in favor of  the formation of the type type $\rm \uppercase\expandafter{\romannumeral2}$b SNe. We study how the internal structure and the nuclearsynthesis are connected with the evolution of the star.
        The main conclusions can be summarized as follow:

   \begin{enumerate}

      \item  SN $\rm \uppercase\expandafter{\romannumeral2}$b show hydrogen lines in early spectra, while later spectra show helium lines but no hydrogen lines. They are believed to originate from the core collapse supernova with very thin hydrogen left in their outer envelope. There are two stripping mechanisms for the formation of type $\rm \uppercase\expandafter{\romannumeral2}$b supernovae. There are strong line-driven winds from isolated massive stars and mass transfer via RLOF in binary systems. Stellar wind in the less massive stars range within $\rm M < 20 M_{\odot}$ is too weak to give rise to type $\rm \uppercase\expandafter{\romannumeral2}$b supernovae because there exists a thick hydrogen envelope. Mass transfer via RLOF provides a promising channel for mass loss that is not solely regulated by the mass loss via stellar winds. Interacting binaries can therefore explain the existence of relatively low-mass progenitors of stripped envelope type $\rm \uppercase\expandafter{\romannumeral2}$b supernova. Some initial parameters, such as rotational velocities, metallicity, overshooting and orbital period, have important impacts on the RLOF and thus the formation of type $\rm \uppercase\expandafter{\romannumeral2}$b Supernova. A larger hydrogen envelope mass indicates a more extended radius and a lower effective temperature, and vice versa.

      \item The faster the initial rotation rate, the greater the mass-loss rate of the stellar winds. Rapid rotation can decrease the low limit of the mass that can turn into a type $\rm \uppercase\expandafter{\romannumeral2}$b supernova due to rotationally enhanced helium cores and stellar winds. Initially, stellar luminosities are lower in rotating models because the effective gravitational acceleration can be reduced by the centrifugal force. Later, stellar luminosity is higher because the helium-core can be enlarged under the influence of Eddington-Sweet circulation. The models with rotation have higher core temperature and lower central density due to more massive convective helium cores. Relatively low-mass helium stars usually experience a rapid expansion of the envelope during the core carbon burning phase. Moreover, the opacity in the radiative envelope can be decreased by the rotational mixing and the corresponding efficiency of the convective dredge-up can be reduced. Rotational mixing can enlarge the main sequence lifetime. Rotational mixing is responsible for the transport of nuclear matter from the core to the surface. Surface chemical species of CNO processing can be changed in rotating models. Surface $\rm ^{4}He$ and $\rm ^{14}N$ are enhanced by rotational mixing while surface $\rm ^{12}C$ and $\rm ^{16}O$ can be decreased. Rotational mixing can affect the hydrogen abundance just outside the helium core, which in turn decreases energy generation from hydrogen shell burning. Rotating star can produce a larger carbon-oxygen core and a higher compactness than non-rotating counterpart.

     \item  The larger convective overshoot is very important for setting the larger size of the convective cores, especially for helium or carbon cores in the advanced evolution of massive stars. It also can increase the stellar luminosity, stellar winds and lifetime in the main sequence. Moreover, overshooting can also decrease the minimum mass for the formation of type $\rm \uppercase\expandafter{\romannumeral2}$b supernovae. The larger convective overshooting develops a larger carbon oxygen core and higher compactness than non-rotating stars. Larger overshooting can restrict the development of the dredge-up and there appears a smaller convective dredge-up region after core hydrogen exhaustion.
         In binary system, the hydrogen burning shell can be extinguished earlier in the model with the larger overshoot. Larger nitrogen enrichments can be reproduced by both RLOF and the subsequent strong WR winds.

     \item Stars with different initial metallicity have different pre-supernova structures. Most importantly, metallicity has an important impact on mass loss via stellar winds. If the amount of mass lost is very low, the helium core and the compactness of the presupernova star will be larger. Moreover, low metallicity implies a smaller initial helium mass fraction and the final helium core mass can be reduced. Low metallicity decreases the energy generation of the hydrogen shell burning via the CNO cycle, and this decreases the boundary for the helium core. The opacity reduces with the decreasing of the metallicity. The combined effects of opacity, energy generation, and mass loss determine whether the progenitor ends up as a red supergiant or a blue supergiant. The primary star with lower metallicity is prone to generate the more compact blue progenitors and can remain less hydrogen mass during RLOF. The more compact radiative structure of the blue supergiant envelope places greater boundary pressure on the helium core and it has important effect on the subsequent evolution. The donor with higher metallicity tends to give rise to higher effective temperature SN $\rm \uppercase\expandafter{\romannumeral2}$b progenitor.

     \item Compared with single stars, the primary stars in binary systems, develop less massive helium and carbon-oxygen cores.
     This is expected by the losing mass due to mass transfer. As hydrogen is converted into helium in the core, the interior of the star becomes progressively less sensitive to variations in the total mass but to the mass of the helium core. Close binary evolution should lead to a further stripping of the hydrogen envelope, and the formation of SN $\rm \uppercase\expandafter{\romannumeral2}$b supernovae is extremely sensitive to the initial orbital period.
     The effective temperature gradually decreases with increasing initial orbital period while the size of the radius goes up. This can be explained by the fact that less hydrogen is eliminated in the case of the binary with wider orbit. The system with the range $\rm  \sim10$ days $\rm < P_{orb}<$ 700 days may turn into type SNe $\rm \uppercase\expandafter{\romannumeral2}$b. The model with $\rm P_{orb}=3.0$ days loses all of their hydrogen-rich envelopes and become SNe Ib whereas the system with initial $\rm P_{orb}>$ 700 days finally explodes as a RSG SNe $\rm \uppercase\expandafter{\romannumeral2}$P. The system with 300 days $\rm <P_{orb}<$ 700 days can give rise to the RSG type SN $\rm \uppercase\expandafter{\romannumeral2}$b progenitor. The initial period of 300 days roughly separates the RSG type SN $\rm \uppercase\expandafter{\romannumeral2}$b progenitor from the YSG type progenitors. The binary system B7 with an initial $\rm P_{orb}=30$ days can produce BSG type SN $\rm \uppercase\expandafter{\romannumeral2}$b progenitor.
     The primary star in the binary system may end its live as a compact BSG and this depends on the larger overshooting parameter and initial orbital period.

     \item  The mass fraction of $\rm ^{12}C$ left in the core when core He is depleted can significantly affect the structure of the progenitor of supernovae. Generally, the mass fraction of $\rm ^{12}C$ is left in the core after core He burning falls down with the increasing of the CO core mass. This can contribute to a faster outward shift of the $\rm ^{12}C$ burning shell and make the core of the star more compact. The fraction of $\rm ^{12}C$ left heavily depends on the mass loss via RLOF, metallicity, overshooting and initial rotational velocity. The remaining mass fraction of $\rm ^{12}C$ is higher in an initial tighter binary system with smaller overshooting, lower initial rotational velocities and metallicity.

   \end{enumerate}

\section*{acknowledgments}
 We are very grateful to an anonymous
referee for his/her valuable suggestions and very insightful
remarks, which have improved this paper greatly. This work was sponsored by the  National Natural Science Foundation
of China (Grant Nos. 11863003, 12173010), Swiss National Science Foundation (project number 200020-172505),  Science and technology plan projects of Guizhou province  (Grant No. [2018]5781). Dr. Y. Qin gratefully acknowledges
the Science Foundation of University in Anhui Province (Grant
No. KJ2021A0106).


\bibliographystyle{aasjournal}
\begin{figure}[h]
  \centering
  \includegraphics[width=0.9\textwidth]{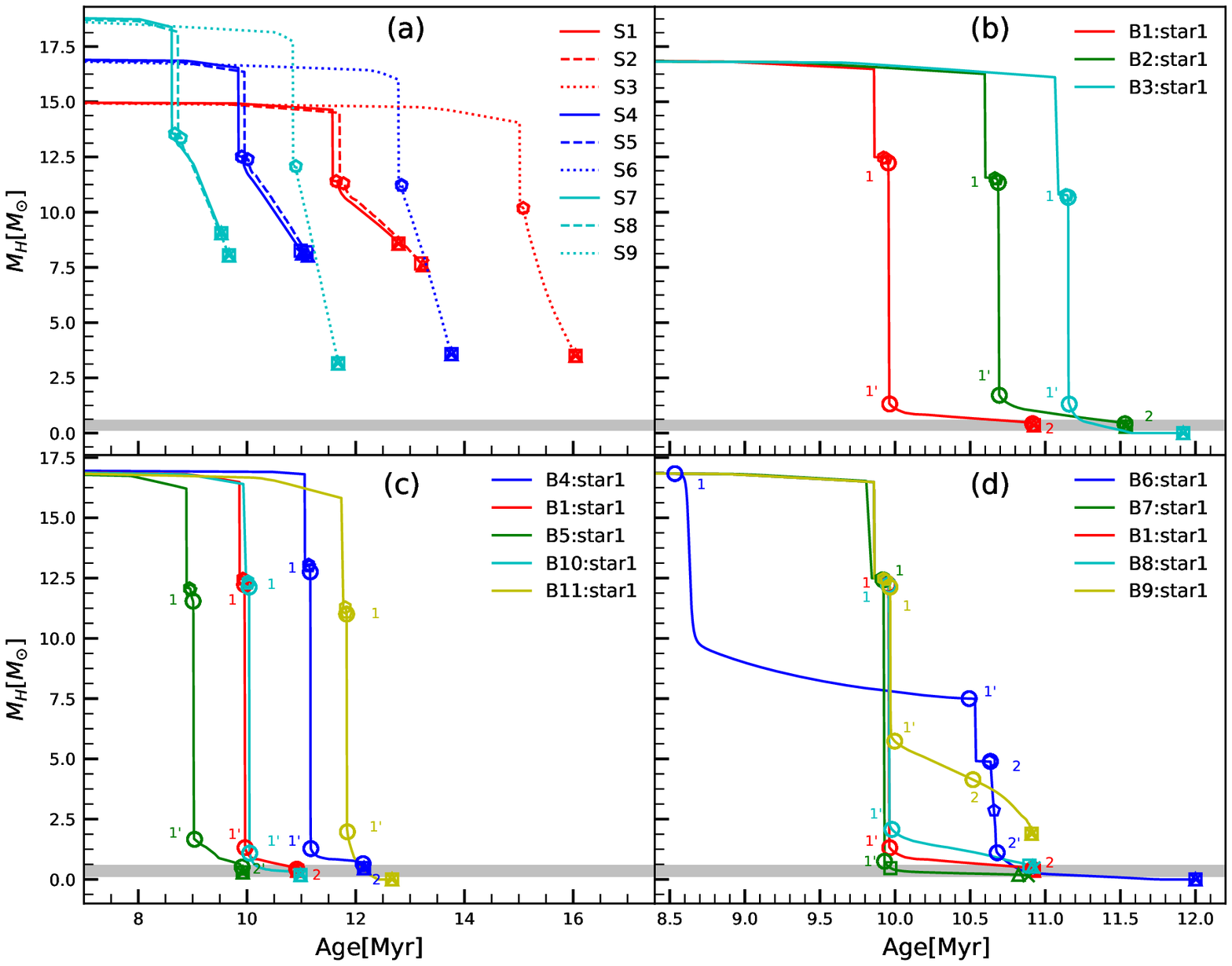}\\
  \caption{The mass of the hydrogen envelope as a function of evolutionary ages. Panel (a): For the single stars with various initial masses and rotational velocities.
  Panel (b): For the primary stars with various overshooting parameters in the binary system with initial $\rm P_{orb}=300$ days. Panel (c): For the primary stars with various initial rotational velocities and metallicities in the binary system with  initial $\rm P_{orb}=300$ days.
  Panel (d): For the primary stars in the binary system with various initial orbital periods. Observations for the hydrogen envelope masses of
  SNe $\rm \uppercase\expandafter{\romannumeral2}$b which are indicated in shaded regions are $\rm M_{H}=0.033- 0.5 M_{\odot}$.\label{fig:general 1}}
\end{figure}

\begin{figure}[h]
  \centering
  \includegraphics[width=0.9\textwidth]{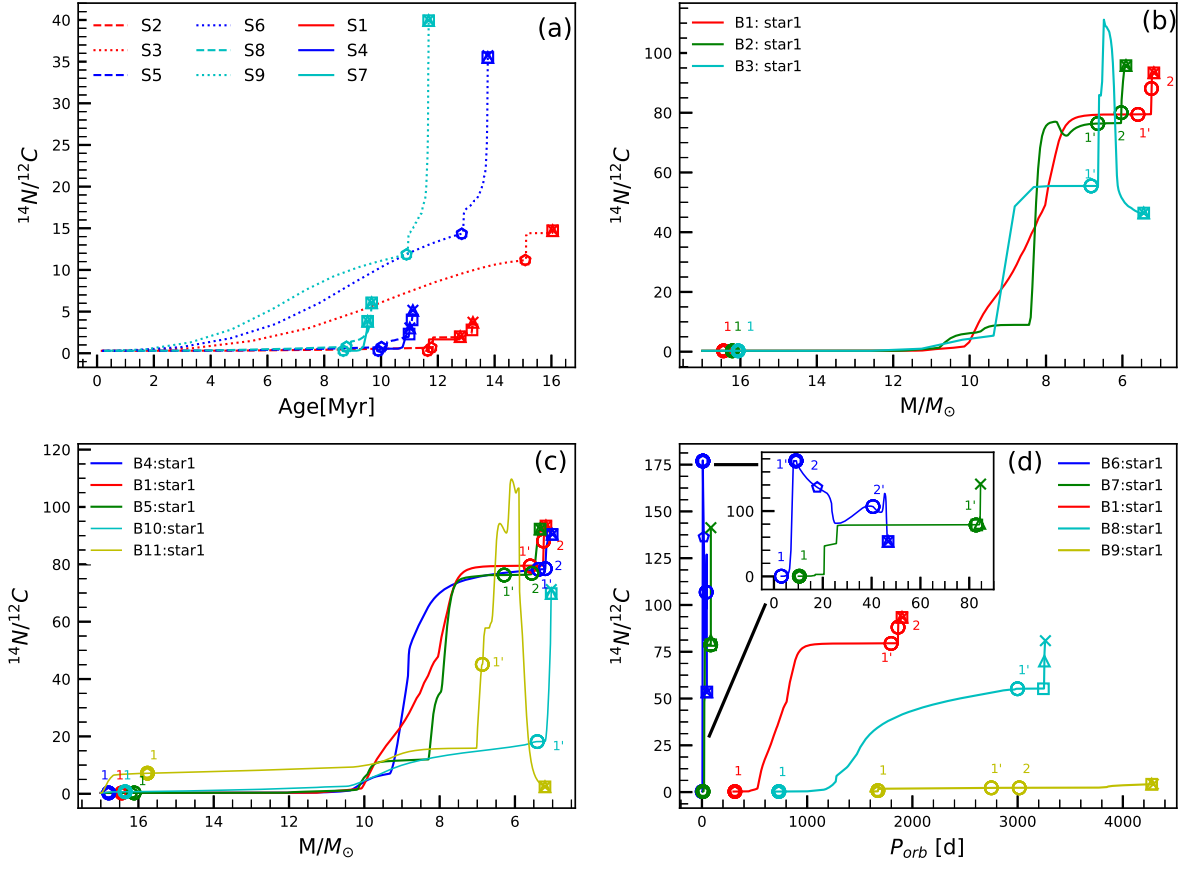}\\
  \caption{Panel (a): Surface mass fraction ratio of nitrogen to carbon as a function of the evolutionary age for the single stars with different initial mass and rotational velocities.
  Panel (b): This ratio as a function of the actual mass for the primary stars with different overshooting parameters in the binary system with an initial $\rm P_{orb}=300$ days. Panel (c): This ratio as a function of the actual mass for the primary stars with different initial metallicities and rotational velocities in the binary system with an initial $\rm P_{orb}=300$ days. Panel (d): This ratio as a function of the orbital period for the primary stars in the binary system with different initial orbital periods.\label{fig:general 2}}
\end{figure}

\begin{figure}[h]
  \centering
  \includegraphics[width=0.9\textwidth]{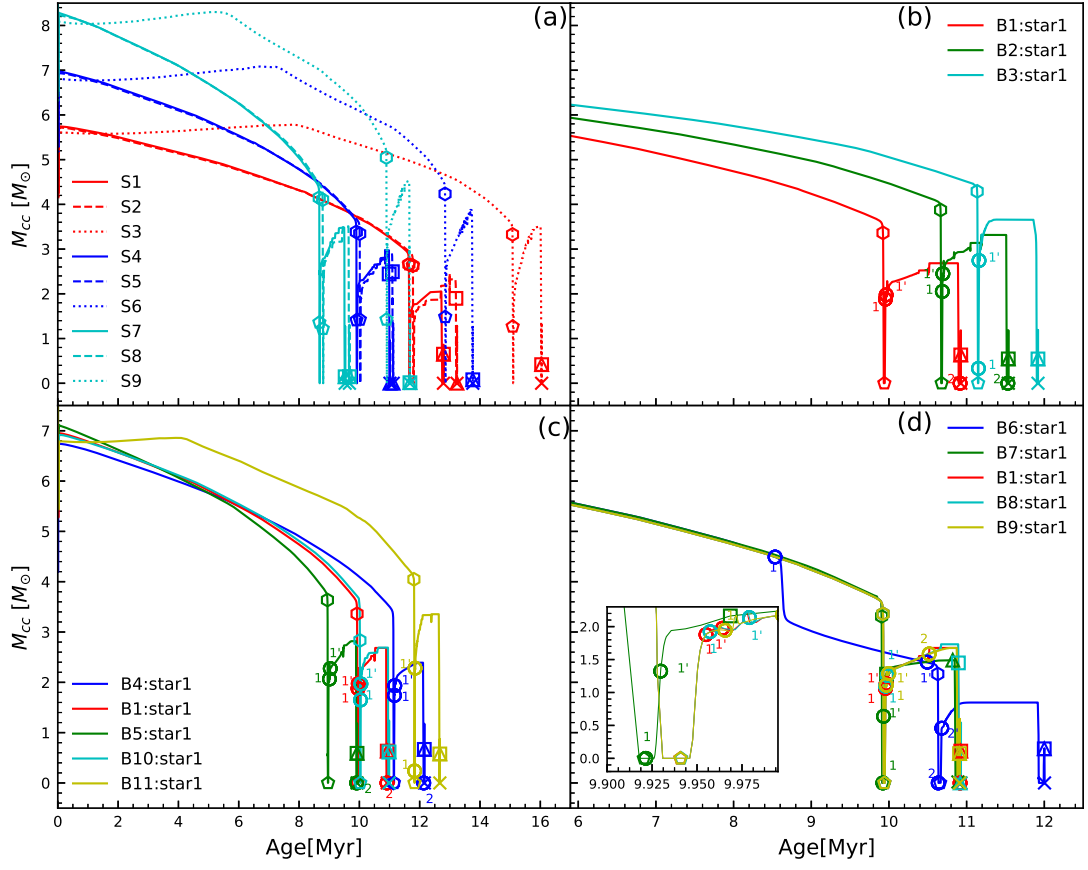}\\
  \caption{ The convective core as a function of evolutionary ages. Panel (a): For the single stars with various initial mass and rotational velocity.
  Panel (b): For the primary stars with various overshooting parameters in the binary system with an initial $\rm P_{orb}=300$ days. Panel (c): For the primary stars with various initial rotational velocities and metallicities in the binary system with an initial $\rm P_{orb}=300$ days. Panel (d): For the primary stars in the binary system with various initial orbital periods.\label{fig:general 3}}
\end{figure}

\begin{figure}[h]
  \centering
  \includegraphics[width=0.9\textwidth]{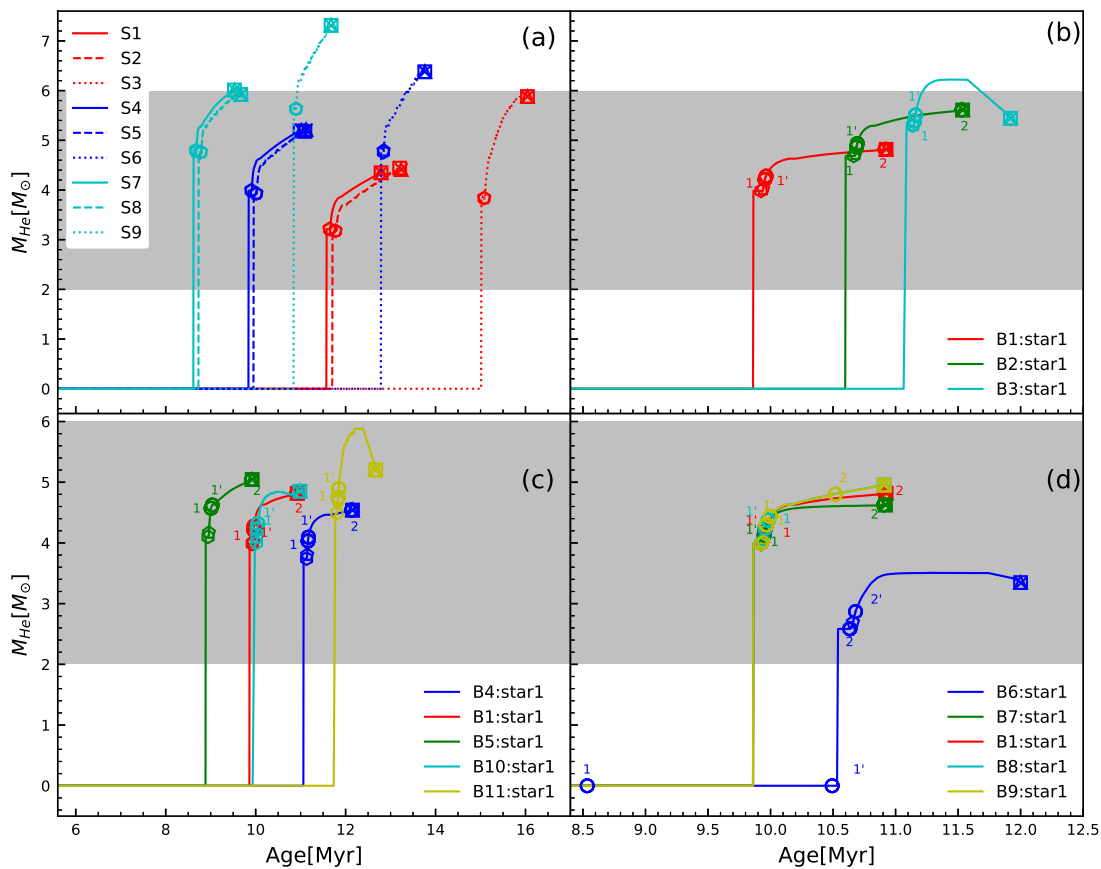}\\
  \caption{The mass of the helium core as a function of the evolutionary age. Panel (a): For the single stars with different initial mass and rotational velocities.
  Panel (b): For the primary stars with different overshooting parameters in the binary system with an initial $\rm P_{orb}=300$ days. Panel (c): For the primary stars with different initial metallicities and rotational velocities in the binary system with an initial $\rm P_{orb}=300$ days. Panel (d): For the primary stars in the binary system with different initial orbital periods. Observations for helium core masses of
  SNe $\rm \uppercase\expandafter{\romannumeral2}$b which are indicated in shaded regions are $\rm 2- 6.0 M_{\odot}$.\label{fig:general 4}}
\end{figure}

\begin{figure}[h]
  \centering
  \includegraphics[width=0.9\textwidth]{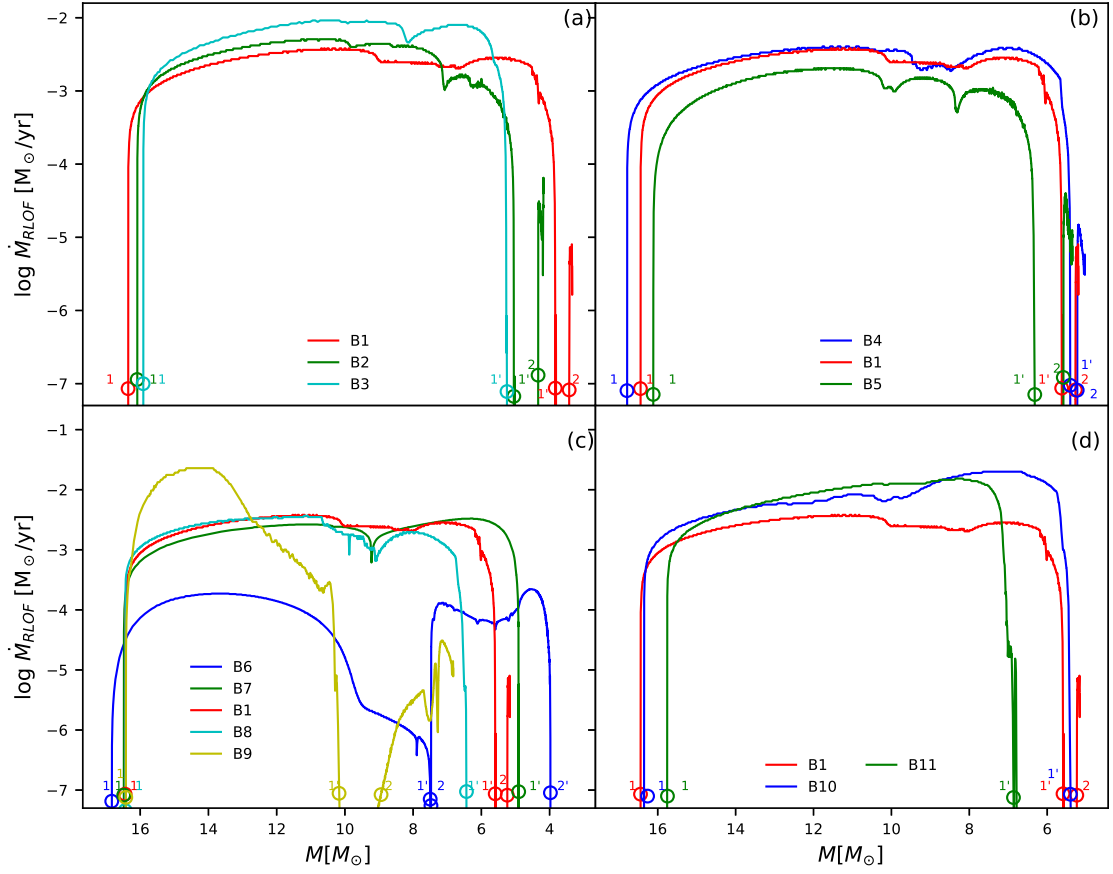}\\
  \caption{The rate of mass transfer via RLOF as a function of evolutionary ages.
  Panel (a): For the primary stars with various overshooting parameters in the binary system with an initial $\rm P_{orb}=300$ days. Panel (b): For the primary stars with various initial metallicities in the binary system with an initial $\rm P_{orb}=300$ days.
  Panel (c): For the primary stars in the binary system with various initial orbital periods. Panel (d): For the primary stars with various initial rotational velocities in the binary system with an initial $\rm P_{orb}=300$ days.\label{fig:general 5}}
\end{figure}

\begin{figure}[h]
  \centering
  \includegraphics[width=0.7\textwidth]{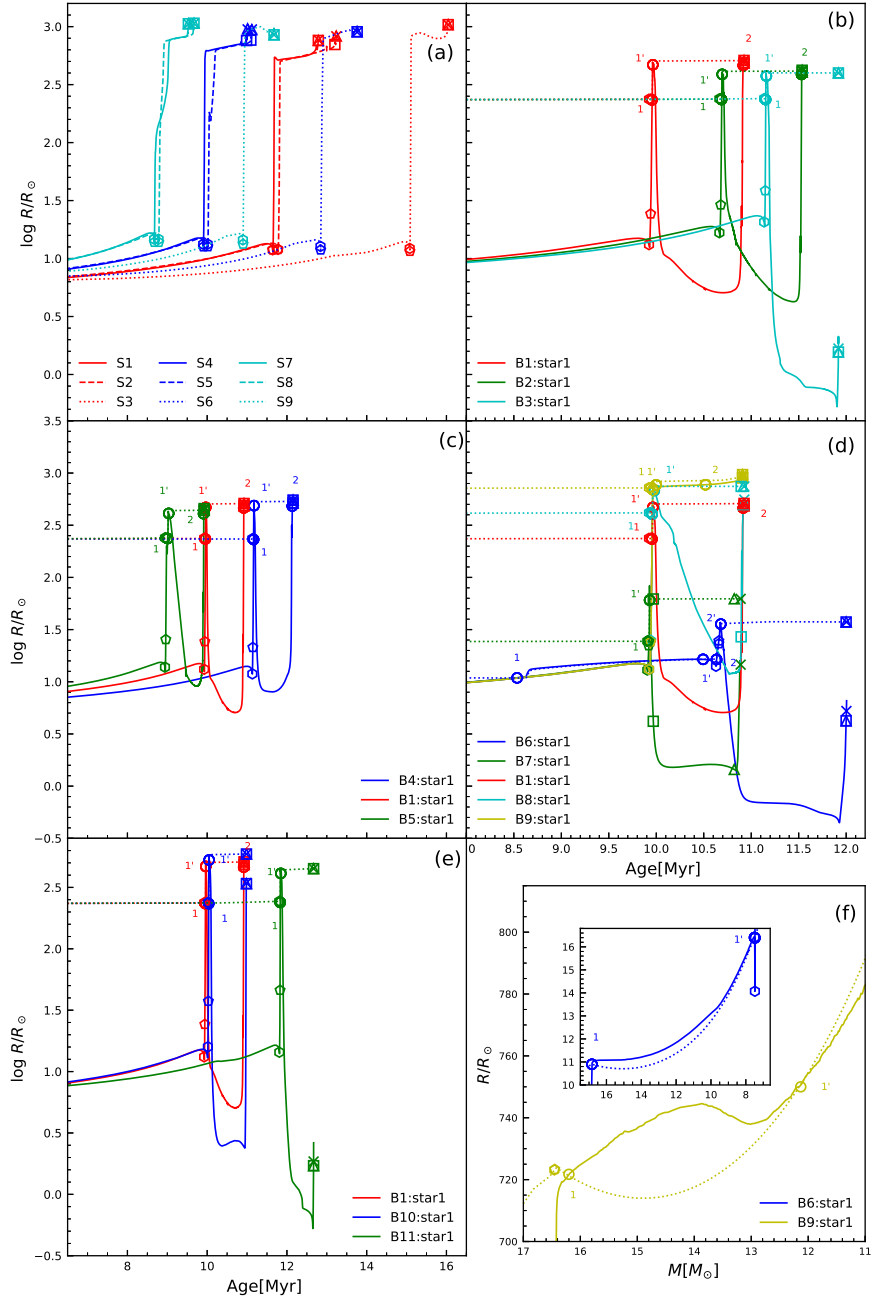}\\
  \caption{The evolution of photospheric radii as a function of evolutionary ages. Panel (a): For the single stars with different initial mass and rotational velocities.
  Panel (b): For the primary stars with different overshooting parameters in the binary system with an initial $\rm P_{orb}=300$ days. Panel (c): For the primary stars with different initial metallicities in the binary system with an initial $\rm P_{orb}=300$ days.
  Panel (d): For the primary stars in the binary system with different initial orbital periods. Panel (e): For the primary stars in the binary system with an initial $\rm P_{orb}=300$ days but with different initial rotational velocities.
  Panel (f): The evolution of photospheric radii as a function of stellar mass in the binary systems B6 with an initial $\rm P_{orb}=300$ days and B9 with an initial $\rm P_{orb}=1600$ days. The dotted lines correspond to the evolution of the corresponding Roche lobe.\label{fig:general 6}}
\end{figure}

\begin{figure}[h]
  \centering
  \includegraphics[width=0.7\textwidth]{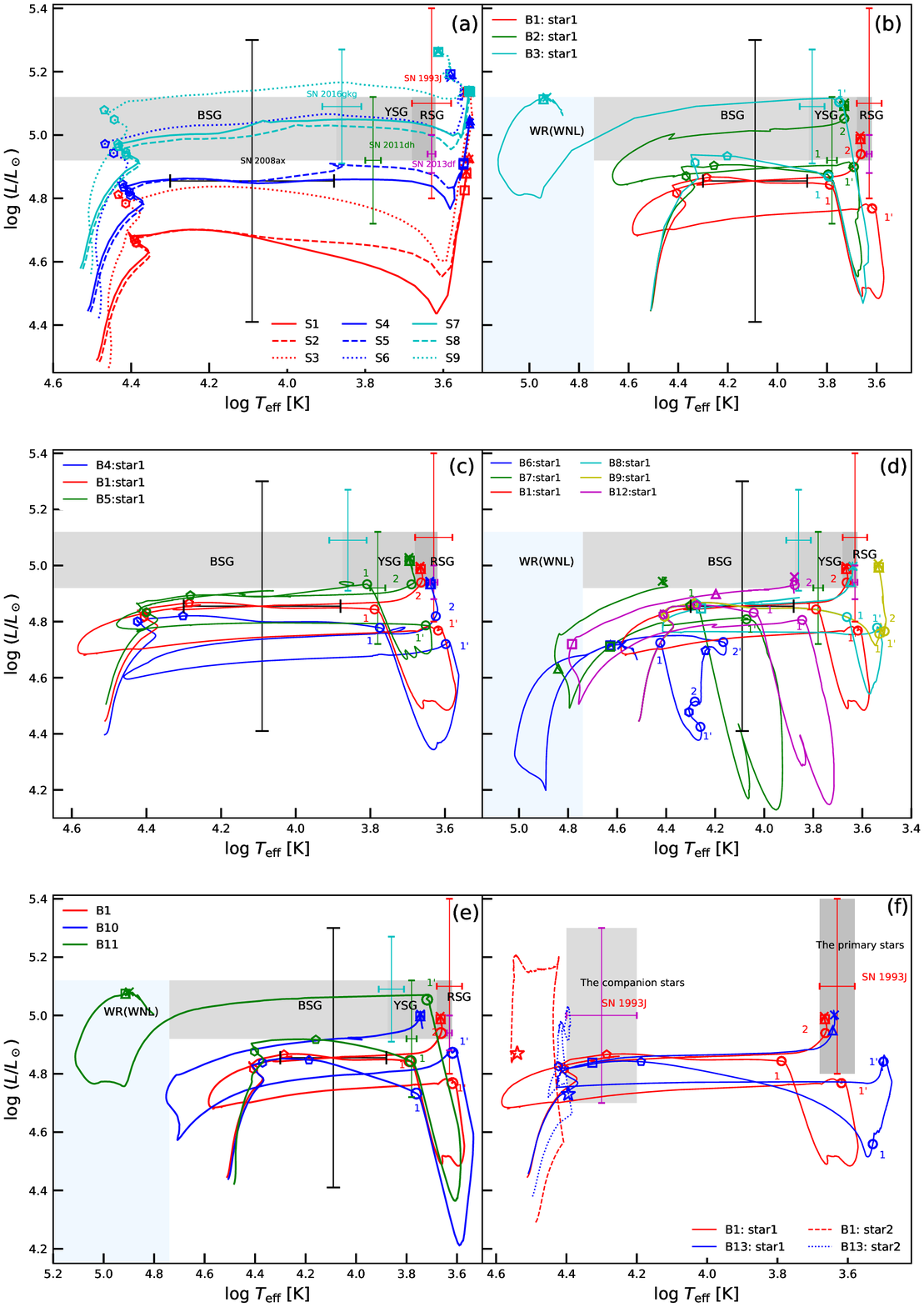}\\
  \caption{Evolutionary tracks of the stars on the Hertzsprung-Russell diagram. Panel (a): For the single stars with different initial mass and rotational velocities. Panel (b): For the primary stars with different overshooting parameters in the binary system with an initial $\rm P_{orb}=300$ days. Panel (c): For the primary stars with different initial metallicities in the binary system with an initial $\rm P_{orb}=300$ days. Panel (d): For the primary stars in the binary system with different initial orbital periods. Panel (e): For the primary stars in the binary system with an initial $\rm P_{orb}=300$ days but with different initial rotational velocities.
  Panel (f): The evolution of the theoretical models B1 and B13. The observational regions of two component stars for the SN 1993J progenitor are marked in the HR diagram.
  Observations for various types of SNe $\rm \uppercase\expandafter{\romannumeral2}$b are indicated in different shaded regions (Gilkis \& Arcavi 2022). Five type $\rm \uppercase\expandafter{\romannumeral2}$b Supernovae are marked in the HR diagram and their observations are from Sravan et al. (2020).\label{fig:general 7}}
\end{figure}


\begin{figure}[h]
  \centering
  \includegraphics[width=0.9\textwidth]{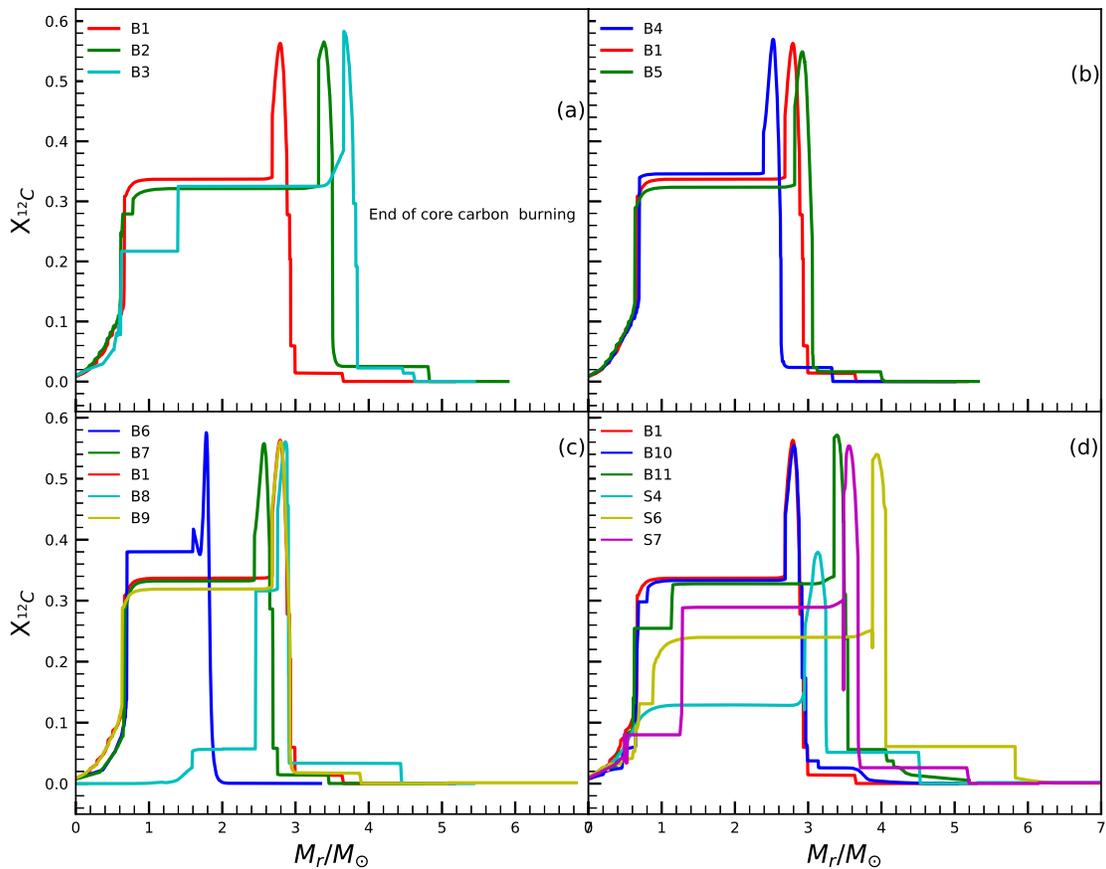}\\
  \caption{Carbon mass fraction as a function of the mass coordinate at the moment of core carbon depletion.
  Panel (a): For the primary stars with different overshooting parameters in the binary system with an initial $\rm P_{orb}=300$ days. Panel (b): For the primary stars with different initial metallicities in the binary system with an initial $\rm P_{orb}=300$ days. Panel (c): For the primary stars in the binary system with different initial orbital periods. Panel (d): For the primary stars with different initial orbital periods and
  for the single stars with different initial mass and rotational velocities.\label{fig:general 8}}
\end{figure}


\end{document}